\documentclass[useAMS,usenatbib,usegraphicx]{mnras}

\def\araa{ARA\&A}%
\def\apj{ApJ}%
\def\apjl{ApJ}%
\def\apjs{ApJS}%
\def\ao{Appl.~Opt.}%
\def\aap{A\&A}%
\def\aapr{A\&A~Rev.}%
\def\mnras{MNRAS}%
\def\pasj{PASJ}%
\def\nat{Nature}%
\def\nphysa{Nucl.~Phys.~A}%
\usepackage{graphicx}
\usepackage[latin1]{inputenc}
\usepackage{color}
\usepackage{times}
\usepackage{natbib}
\usepackage{setspace}
\usepackage{hyperref}
\newif\ifAMStwofonts
\AMStwofontstrue
\definecolor{red}{rgb}{1,0.,0.}

\newcommand{\gaea}{\sc{gaea}}

\newcommand{\msun}{{\rm M}_\odot}
\newcommand{\msunyr}{{\rm M}_\odot\ {\rm yr}^{-1}}
\def\lesssim{\lower.5ex\hbox{$\; \buildrel < \over \sim \;$}}
\def\gtrsim{\lower.5ex\hbox{$\; \buildrel > \over \sim \;$}}
\title[MZRs in {\gaea}] {The evolution of the mass-metallicity
  relations from the VANDELS survey and the {\gaea} Semi-Analytic
  model.} \author[Fontanot et al.]{
  \parbox[t]{\textwidth}{Fabio Fontanot$^{1,2}$\thanks{E-mail:
      fabio.fontanot@inaf.it}, Antonello Calabr\`o$^3$, Margherita
    Talia$^{4,5}$, Filippo Mannucci$^6$, Marco Castellano$^3$,
    Giovanni Cresci$^6$, Gabriella De Lucia$^1$, Anna Gallazzi$^6$,
    Michaela Hirschmann$^7$, Laura Pentericci$^3$, Lizhi Xie$^8$,
    Ricardo Amorin$^{10,11}$, Micol Bolzonella$^5$, Angela
    Bongiorno$^3$, Olga Cucciati$^5$, Fergus Cullen$^{12}$,
    Johan~P.~U. Fynbo$^{13}$, Nimish Hathi$^{14}$, Pascale
    Hibon$^{15}$, Ross~J. McLure$^{12}$, Lucia Pozzetti$^5$}
    \vspace*{8pt}\\
    $^1$ INAF - Astronomical Observatory of Trieste, via G.B. Tiepolo 11, I-34143 Trieste, Italy \\
    $^2$ IFPU - Institute for Fundamental Physics of the Universe, via Beirut 2, 34151, Trieste, Italy \\
    $^3$ INAF - Astronomical Observatory of Rome, Via Frascati 33, I-00040 Monte Porzio Catone (RM), Italy \\
    $^4$ Dipartimento di Fisica e Astronomia, Universit\'a di Bologna, Via Gobetti 93/2, I-40129, Bologna, Italy \\
    $^5$ INAF - Astrophysics and Space Science Observatory of Bologna, via P. Gobetti 93/3,I-40129, Bologna, Italy \\
    $^6$ INAF - Astrophysical Observatory of Arcetri, Largo E. Fermi 5, I-50125, Firenze, Italy \\
    $^7$ DARK, Niels Bohr Institute, University of Copenhagen, Lyngbyvej 2, DK-2100 Copenhagen, Denmark \\
    $^8$ Tianjin Astrophysics Center, Tianjin Normal University, Binshuixidao 393, 300384, Tianjin, China\\
    $^{10}$ Instituto de Investigaci\'on Multidisciplinar en Ciencia y Tecnolog\'ia, Universidad de La Serena, Raul Bitr\'an 1305, La Serena, Chile \\
    $^{11}$ Departamento de F\'isica y Astronom\'ia, Universidad de La Serena, Av. Juan Cisternas 1200 Norte, La Serena, Chile \\
    $^{12}$ SUPA Scottish Universities Physics Alliance, Institute for Astronomy, University of Edinburgh, Royal Observatory, Edinburgh EH9 3HJ \\
    $^{13}$ The Cosmic Dawn Center, Niels Bohr Institute, University of Copenhagen, Juliane Maries Vej 30, DK-2100 Copenhagen, Denmark \\
    $^{14}$ Space Telescope Science Institute, 3700 San Martin Drive, Baltimore, MD 21218, USA\\
    $^{15}$ ESO-Chile, Alonso de Cordova 3107, Vitacura, Santiago, Chile \\
}

\begin{document}
\date{Accepted ... Received ...}

\maketitle

\begin{abstract} 
In this work, we study the evolution of the mass-metallicity relations
(MZRs) as predicted by the GAlaxy Evolution and Assembly ({\gaea})
semi-analytic model. We contrast these predictions with 
recent results from the VANDELS survey, that allows us to expand the
accessible redshift range for the stellar MZR up to $z\sim3.5$. We
complement our study by considering the evolution of the gas-phase MZR
in the same redshift range. We show that {\gaea} is able to reproduce
the observed evolution of the $z<3.5$ gas-phase MZR and $z<0.7$
stellar MZR, while it overpredicts the stellar metallicity at
$z\sim3.5$. Furthermore, {\gaea} also reproduces the so-called
fundamental metallicity relation (FMR) between gas-phase metallicity,
stellar mass and star formation rate (SFR). In particular, the
gas-phase FMR in {\gaea} is already in place at $z\sim5$ and shows
almost no evolution at lower redshift. {\gaea} predicts the existence
of a stellar FMR, that is, however, characterized by a relevant
redshift evolution, although its shape follows closely the gas-phase
FMR. We also report additional unsolved tensions between model and
data: the overall normalization of the predicted MZR agrees with
observations only within $\sim$0.1 dex; the largest discrepancies are
seen at $z\sim3.5$ where models tend to slightly overpredict observed
metallicities; the slope of the predicted MZR at fixed SFR is too
steep below a few $\msunyr$. Finally, we provide model predictions for
the evolution of the MZRs at higher redshifts, that would be useful in
the context of future surveys, like those that will be performed with
JWST.
\end{abstract}


\vspace*{8pt}

\section{Introduction}\label{sec:intro}                     
The observed scaling relations between galaxy properties have always
been considered primary indicators to study the evolution of galaxies,
and constrain the underlying physical processes. As such, the correct
prediction of these relations has always been a key target for
theoretical models of galaxy formation and evolution. The intrinsic
complexity of the non-linear mechanisms acting on the baryonic
component of the Universe prevents an accurate description of the
evolution of the baryonic components of the Large Scale Structure,
whose statistical properties are well recovered using numerical
techniques (i.e. N-body simulations).

The study of the relative content of chemical elements in galaxies of
different properties provides, in particular, tight constraints on
theoretical models. Metals, i.e. chemical elements other than Hydrogen
and Helium, are synthesised in the different stages of stellar
evolution, corresponding to the different nuclear burning sequences
that power stellar emission. As such, the stellar metallicity of a
given galaxy, i.e. its content of metals relative to Hydrogen and
Helium, records the effect of different physical mechanisms, primarily
star formation, but also the balancing between inflows (bringing cold
pristine gas into the system) and outflows (ejecting enriched gas in
the surrounding environment). Moreover, the relative abundances of
specific elements provide insight on the details of the star formation
history, as their production depends on different stellar populations
and different timescales \citep[see][for a
  review]{MaiolinoMannucci19}.

The existence of well defined scaling relations between galaxy
metallicity and stellar mass (mass-metallicity relations or MZR) is
therefore of great interest for understanding galaxy evolution. These
scaling relations hold for both the stellar metallicity ($Z_\star$)
and for the metallicity of the ionized gas ($Z_{\rm g}$, computed from
the $[O/H]$ abundance ratio) in the inter-stellar medium (ISM). The
Sloan Digital Sky Survey (SDSS) spectroscopic observations have
provided a benchmark for the local {\it stellar} MZR
\citep{Gallazzi05} and {\it gas-phase} MZR \citep{Tremonti04}. Further
work showed that local quiescent and star forming galaxies follow two
separate MZRs \citep{Peng15, Trussler20}, that is usually interpreted
as an evidence for ``strangulation'' processes (i.e. the suppression
of star formation activity due to the lack of newly accreted gas onto
the galaxy). As an alternative explanation, \citet{Spitoni17} proposed
that these differences in chemical composition can be explained by
short gas infall timescales in the early phases of formation of
present-day quiescent galaxies and by strong outflows in low-mass star
forming galaxies. No clear evidence for a direct dependence of the MZR
on environment has been found \citep{Pasquali12, Namiki19}.

The stellar MZR shows little evolution at intermediate redshifts
$z\lesssim1$ \citep[e.g.][]{Ferreras09, Gallazzi14}, but a relevant
evolution at higher redshifts \citet{Sommariva12, Cullen19}. A more
significant evolution has been found for the gas-phase MZR at
$1.5<z<3.5$ \citep{Erb06, Maiolino08, PerezMontero13, Zahid14cos,
  Troncoso14, Steidel14, Onodera16, Curti20z}. At all redshifts, the
MZR shows a similar shape, with increasing metallicity at increasing
stellar mass, and an overall metallicity decrease at increasing
redshift. However, the differential evolution is characterized by the
so-called {\it downsizing in metallicity} \citep{Maiolino08,
  Fontanot09b}: low-mass galaxies exhibit a larger metallicity
evolution from $z\sim3.5$ to $z\sim0$ than their more massive
counterparts.

Already \citet{Tremonti04} studied secondary dependences in the MZR;
later \citet[see also~\citealt{LaraLopez10, Hunt12}]{Mannucci10}
proposed a fundamental metallicity relation (FMR) between gas-phase
metallicity, stellar mass and star formation rate (SFR). Further
studies at higher redshift suggest a negligible evolution of the FMR
up to $z\sim2.5$ \citep{Cresci12, Cresci19, Sanders20, Curti20z}. This
stability of the FMR is usually interpreted in the framework of
equilibrium models in which gas infall is balanced by star formation
and feedback-driven outflows \citep[see e.g.][]{Dayal13}. The observed
MZRs thus represent projections of the FMR, and the observed redshift
evolution is due to different regions of the FMR being sampled at
different redshifts, because of the evolution of the cosmic star
formation rate. More recent works find that a stronger secondary
dependence is measured with the gas content rather than SFR
\citep{Bothwell13}, and that the SFR-defined FMR is a projection of
this more fundamental relation \citep{Bothwell16, Brown18}. The
accepted interpretation of available data is that the role played by
outflows as SFR regulators is minimal, and that the shape and
evolution of the FMR is mainly driven by the cosmological evolution of
the cold gas content. A discussion about the role of gas (and more
precisely the gas-to-stellar mass ratio) can be found in
\citet{Zahid14a} and \citet{Curti20z}, who suggest that the evolution
of the MZR and FMR is driven by the evolution of the gas/star mass
ratio. Indeed, \citet{Ellison08} suggest that the MZR sensitivity to
SFR efficiency is the most likely origin for the trends they report as
a function of specific star formation (sSFR $= SFR/\msun$) and galaxy
size.

Despite a large amount of focused work, these studies are still
potentially affected by significant biases, mostly due to the
different techniques used to derive metallicities from spectroscopic
observations and to uncertainties on the relative calibrations
\citep[see][for an extensive reviews]{KewleyEllison08,
  MaiolinoMannucci19}. Stellar metallicity estimates mostly rely on
the comparison between stellar absorption features with distinct
sensitivity to age and metallicity (e.g. the Lick indices) and the
predictions from Stellar Population Synthesis (SPS) models. The
derived estimates depend on the SPS models used, on the features or
spectral range analysed, and on the assumptions made for the galaxy
star formation history. Spectral features in the Ultra-Violet (UV) are
also used to estimate the stellar metallicity, mostly in high-redshift
galaxies, of the younger stellar component. Gas metallicities are
usually based on the analysis of strong emission line ratios, that
could be calibrated in two different ways. The so-called direct
approach \citep[see e.g.][]{Pilyugin10, PerezMontero14} relies on
measuring the temperature of the different ionization
zones. Alternatively photoionization models are used to generate a
grid of synthetic spectra to be compared with observations.

It is important to keep in mind that most of these approaches need an
absolute calibration that is usually obtained by comparison with
theoretical models of stellar evolution. Moreover, HII regions in
galaxies are expected to cover a wide range of physical properties and
direct methods can only give an estimate of the mean emission. These
considerations imply that the accuracy of the measurements is limited
by our knowledge of the physics of ionized regions and/or stellar
evolution and relies on a number of assumptions to break the complex
degeneracy between age and metallicity. These considerations explain
the apparent tensions between results based on different
methods. Photoionization models tend to overestimate the gas-phase
metallicity by 0.2-0.6 dex \citep{KewleyEllison08}, while the direct
methods may differ by up to a factor of a few
\citep{Tsamis03}. Generally, direct measurement using UV spectra
provide a good match of the metallicities of young stellar population,
although they might be biased toward lower metallicities due to the
presence of temperature fluctuations \citep{Bresolin16, Curti20a}.

Theoretical models of galaxy evolution have long been reported to face
tensions in reproducing the observed MZRs \citep{SomervilleDave15},
and in particular their redshift evolution. These discrepancies have
been reduced with the new generation of models \citep{Hirschmann16,
  Xie17, Lagos16, Torrey19}: these models also provide a better
overall description of the assembly history of the galaxy populations,
solving long standing issues with the evolution of low-mass galaxies
(both semi-analytic - \citealt{Fontanot09b} - and numerical
simulations \citealt{Weinmann12}) and the local fraction of quenched
galaxies \citep{DeLucia19, Xie20}.

In a recent series of papers, we introduce the GAlaxy EVolution and
Assembly ({\gaea}) model. This model is able to correctly reproduce
the evolution of the GSMF and cosmic SFR up to the highest redshift
accessible \citep{Fontanot17b}. The same model is also able to
reproduce both the stellar and gas-phase $z\sim0$ MZRs
\citep{Hirschmann16}. In particular, \citet{DeLucia20} study the
scatter around the $z\sim0$ gas-phase MZR, showing that it is
primarily regulated by the cold gas inflow/accretion rate on the model
galaxies. In this paper we now focus on the redshift evolution of the
MZRs and FMR as predicted by {\gaea}, going beyond the preliminary
comparison shown in \citep{Hirschmann16}.

This paper is organised as follows. In Section~\ref{sec:gaea}
and~\ref{sec:data} we will present, respectively, our semi-analytic
model of galaxy formation and the datasets we will consider for the
comparison with the theoretical predictions shown in
Section~\ref{sec:results}. We will then discuss our conclusions in
Section~\ref{sec:discussion}. Finally, we will summarise our
conclusions in Section~\ref{sec:final}.

\section{Semi-analytic model}\label{sec:gaea}
In this work we consider predictions from the GAlaxy Evolution and
Assembly ({\gaea}) semi-analytic model (SAMs). SAMs represent a
theoretical tool to predict the evolution of galaxy populations across
cosmological epochs and volumes. They assume that galaxies form inside
dark matter haloes from the condensation of gas. A complex network of
physical processes is then responsible for the cooling and heating of
the gas, as well as for the matter and energy exchanges between the
different gas phases and galaxy components (disc, bulge and halo). In
order to amend for our limited knowledge of the details of the
relevant mechanisms, SAMs are defined by using a system of
differential equations based on parametrizations constructed either on
empirical results or theoretical arguments. This approach translates
into a limited computational request (when compared, e.g., with
hydrodynamical simulations) and thus allows a characterization of the
role of individual mechanisms in shaping galaxy properties.

{\gaea} is especially well suited for the comparison presented in this
paper, thanks to its improved modelling of stellar feedback
\citep{Hirschmann16}. This leads to a prediction for the evolution of
the cosmic star formation rate and galaxy stellar mass function in
good agreement with observational constraints up to the highest
redshifts available \citep{Fontanot17b}. In particular, {\gaea} is
able to reproduce the evolution of the low-mass end of the galaxy
stellar mass function at $z<3$, that has represented a long standing
problem for models of galaxy evolution \citep{Fontanot09b,
  Weinmann12}. This success is due to the implementation of an
ejective feedback scheme that combines (a) the numerical results for
gas reheating from \citet{Muratov15}, (b) the outflow rate description
from \citet{Guo11} and (c) the gas reincorporation analysis from
\citet{Henriques13}. Given the correct distribution of stellar masses
at different cosmic epochs in {\gaea}, checking the chemical enrichment
levels is thus an interesting test for this SAM. We already show in
\citet{Hirschmann16} that {\gaea} is able to reproduce the evolution
of the MZRs up to $z\sim2$. In this paper, we will deepen this
analysis using the most recent results covering the redshift range
$0.7<z<3.5$ both for the stellar and gas-phase MZRs and FMR.

{\gaea} also includes an advanced treatment for chemical enrichment
\citep{DeLucia14}, that allows us to trace the evolution of the
abundances of different metal species. This scheme accounts for the
different lifetime of stars of different initial mass
\citep{PadovaniMatteucci93} and traces their individual contribution
to chemical enrichment, via differential yields. In detail, we assume
the following yields: (a) for the lowest mass stars contributing to
the chemical enrichment of the ISM (i.e. $m_\star<8 \msun$, that end
their lives as asymptotic giant branch stars) we consider the
\citet{Karakas10} yields; (b) more massive stars are expected to end
their evolution as SNeII, whose yields are taken from
\citet{ChieffiLimongi02}; (c) finally for SNeIa we use yields from
\citet{Thielemann03pap}. These assumptions imply that the global metal
enrichment is a direct {\gaea} prediction and it is not regulated by a
free parameter as is typically the case when adopting an instantaneous
recycling approximation. 

In this paper, we use a sample including all model galaxies (i.e. both
centrals and satellites) extracted from our reference {\gaea}
realization run on the Millennium Simulation ($\Omega_\Lambda=0.75$,
$\Omega_m=0.25$, $\Omega_b=0.045$, $n=1$, $\sigma_8=0.9$, $H_0=73 \,
{\rm km/s/Mpc}$ \citealt{Springel05}). Only for the gas-phase MZR, we
restrict the sample to model galaxies with a meaningful gas fraction
(e.g. $f_{\rm gas}= M_{\rm gas}/M_{\star} >0.1$), as the observational
samples used for comparison are typically composed of
star-forming/gas-rich galaxies. Moreover, this selection removes model
galaxies with unrealistic metallicity levels due to the dearth of
gas. We check that using other selection criteria, such as restricting
the sample to either central galaxies\footnote{The evolution of
  satellite galaxies in {\gaea} depends on the treatment of
  environmental processes \citep[see e.g.][]{DeLucia19}. In the H16F
  model, in particular, satellites are subject to instantaneous
  removal of their hot gas (i.e. ``strangulation''), which affects
  heavily their subsequent evolution.}, or actively star-forming
galaxies (e.g. using a standard $sSFR > 0.3/t_{\rm hubble}$
threshold), does not change our conclusions qualitatively. In all
plots we show the median MZR and FMR, and the corresponding 15th-85th
percentile range.

In the following, gas-phase metallicities have been estimated from the
predicted mass fraction of the oxygen element in the cold-gas phase of
the model galaxy. Stellar metallicities have been computed by dividing
the mass in metals locked into long-lived stars by the total stellar
mass of the model galaxy, that represents the intrinsic metallicity of
the stellar component in {\gaea}. This estimate could differ from
observational estimates used in this work which are based on either
optical absorption features (tracing both Fe and $\alpha$-elements,
weighing more the old stellar populations) or the properties of the
FUV spectral range (tracing mostly Fe, weighing more dusty young
stellar populations). As a general example, we quote the results
discussed in \citet{Nelson18a}, warning the reader that these cannot
be extrapolated to all theoretical models of galaxy evolution. These
authors compare the intrinsic stellar metallicities in the {\sc
  IllustrisTNG} simulation \citep{Pillepich18}, with the corresponding
quantities derived from synthetic mock spectra build using {\sc
  IllustrisTNG} metallicity and star formation histories. For their
model galaxis, they find that spectroscopically derived median
$Z_\star$ are offset low by $\sim$0.2-0.5 dex at $M_{\star} <
10^{10.5} M_\odot$, with respect to intrinsic metallicities.
\begin{figure*}
  \centerline{
    \includegraphics[width=9cm]{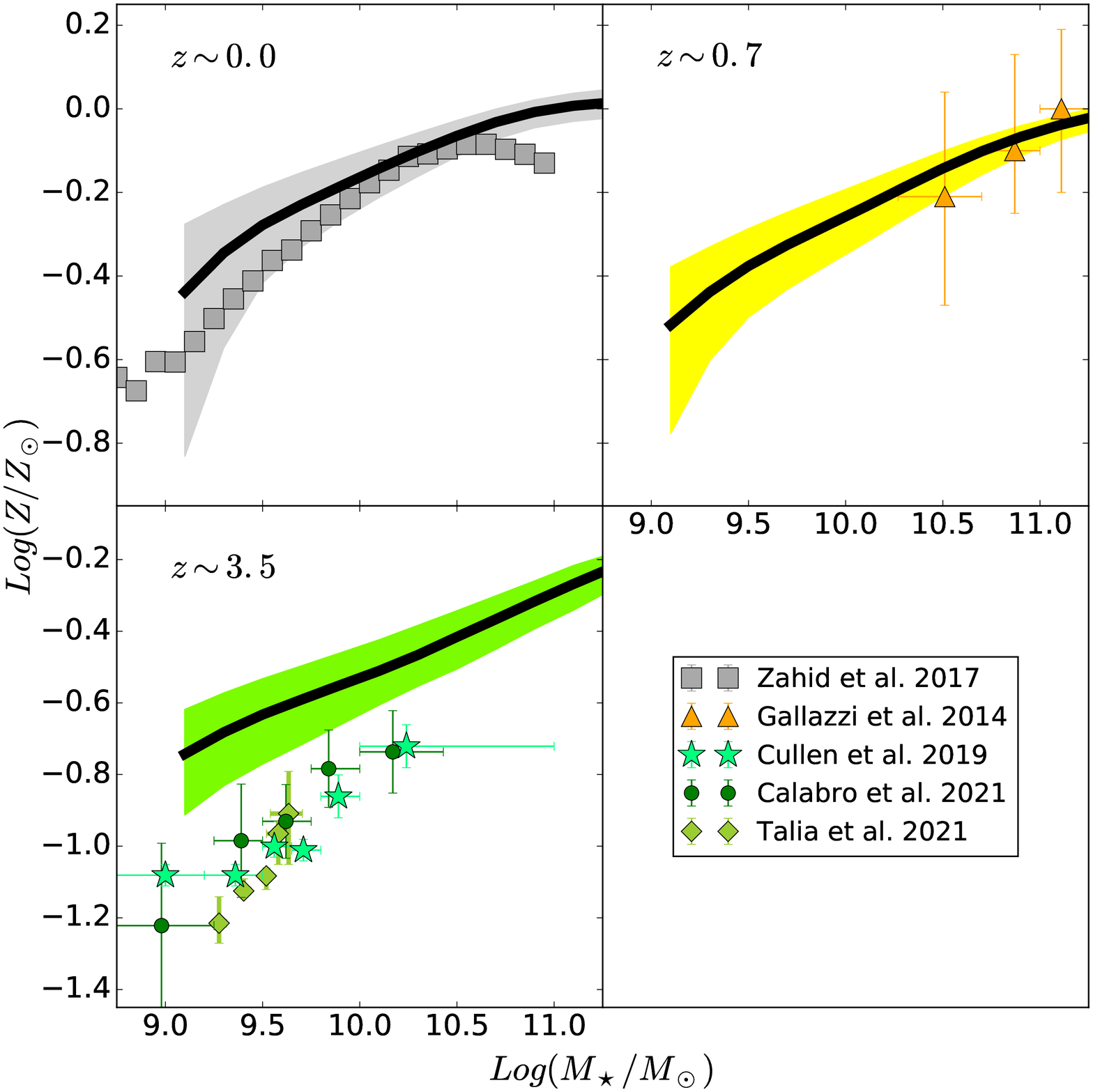}
    \includegraphics[width=9cm]{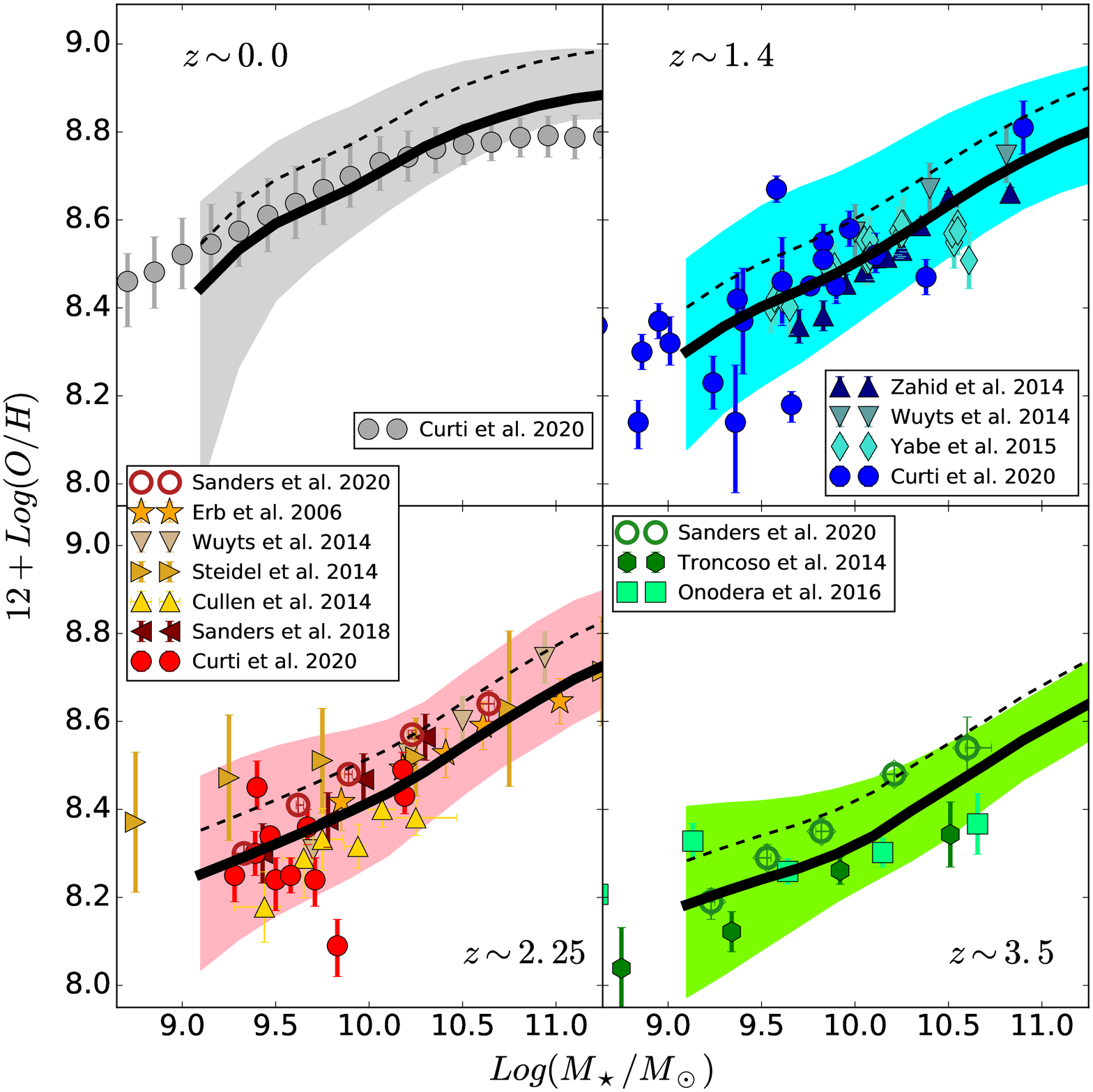} }
  \caption{{\it Left panel}: Redshift evolution of the stellar MZR
    relation. Datapoints with errorbars refer to the estimates from
    \citet[][grey squares]{Zahid17}, \citet[][dark yellow
      triangles]{Gallazzi14}, \citet[][pale green stars]{Cullen19},
    \citet[][green circles]{Calabro20} and \citet[olive
      diamonds]{Talia20}. All data have been rescaled to a common 0.02
    solar metallicity. {\it Right panel}: Redshift evolution of the
    cold gas MZR. Datapoints with errorbars refer to the estimates
    from \citet[][grey circles]{Curti20a}, \citet[light blue and pale
      brown triangles]{Wuyts14}, \citet[][dark blue
      triangles]{Zahid14cos}, \citet[cyan diamonds]{Yabe15},
    \citet[][blue and red circles]{Curti20z}, \citet[orange
      stars]{Erb06}, \citet[orange triangles]{Steidel14}, \citet[gold
      triangles]{Cullen14}, \citet[brown triangles]{Sanders18},
    \citet[green triangles]{Troncoso14} and \citet[open
      circles]{Sanders20}. Solid symbols show data that have been
    rescaled to the same assumptions as in \citet{Curti20z}. In all
    panels, the thick black solid lines represent the corresponding
    median MZRs as predicted by {\gaea} at the appropriate redshift,
    with the shaded area referring to the 15th-85th
    percentiles. Stellar phase MZRs are drawn from a model sample
    including all model galaxies, while gas-phase from a model sample
    including only gas rich galaxies. In the right panel, solid black
    lines refer to {\gaea} predictions shifted 0.1 dex downwards ,
    while the thin dashed lines represent the intrinsic model
    predictions (see text for more details).}\label{fig:mzr}
\end{figure*}
\begin{figure*}
  \centerline{ \includegraphics[width=9cm]{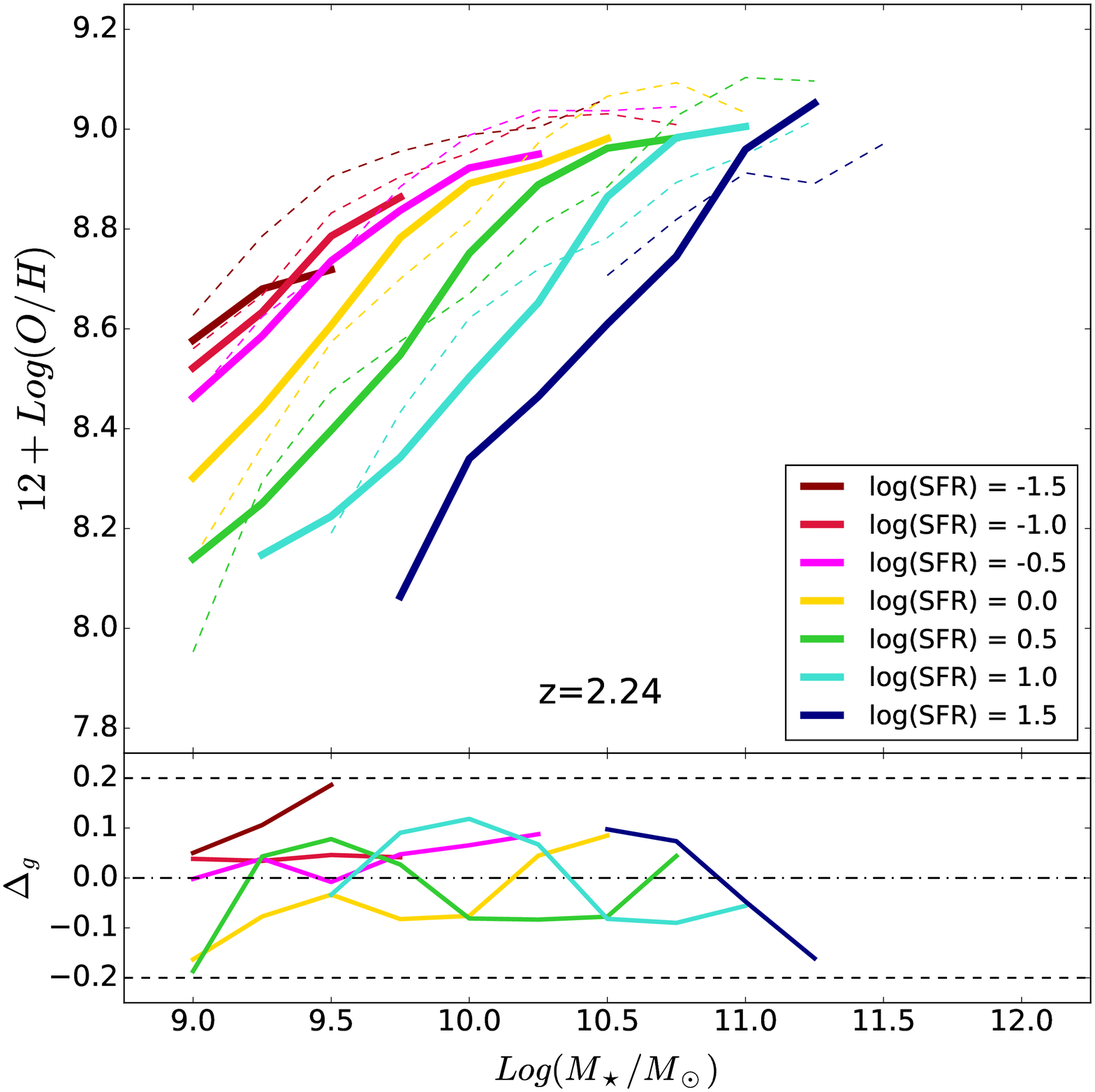} 
    \includegraphics[width=9cm]{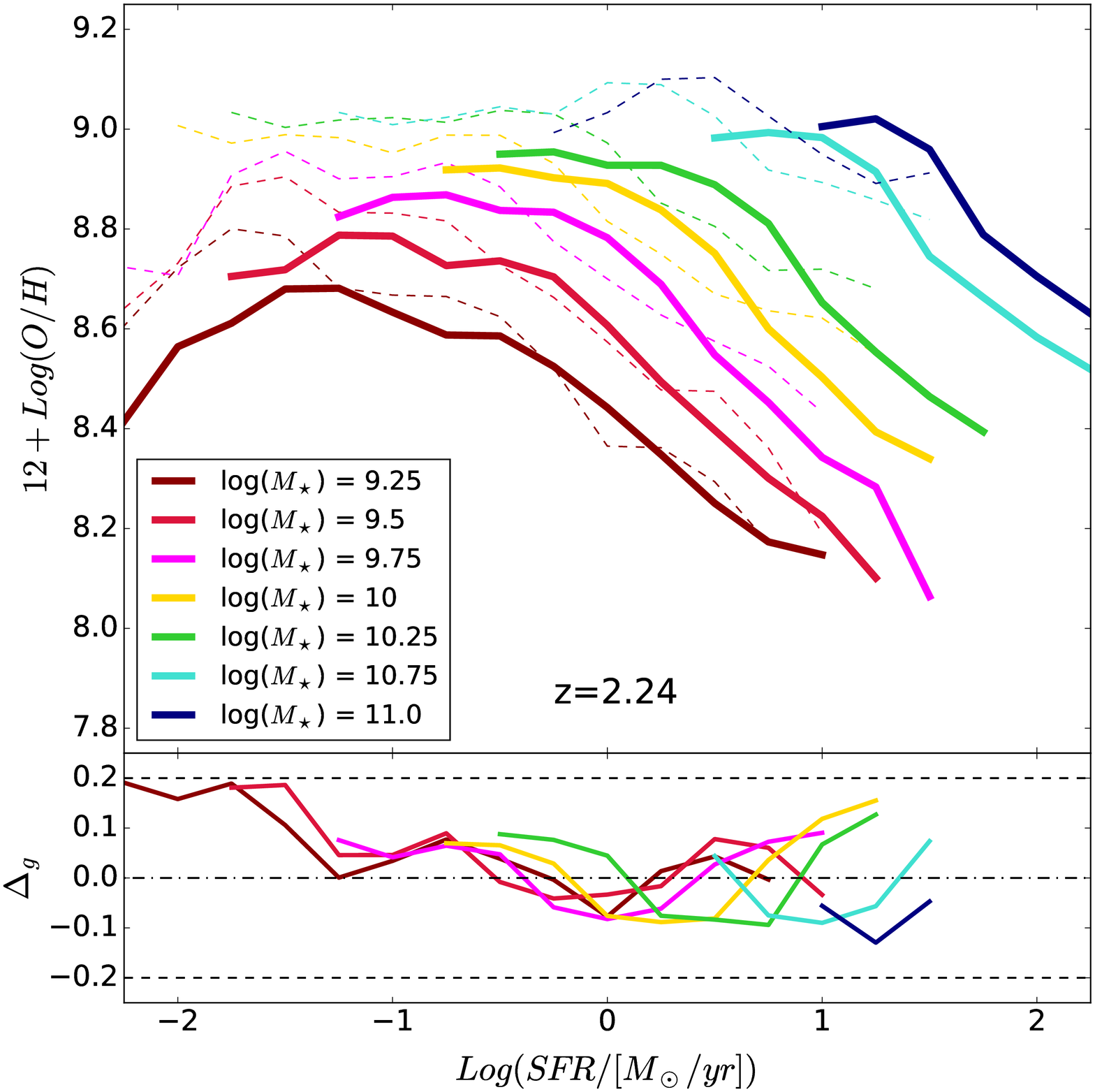} }
  \centerline{ \includegraphics[width=9cm]{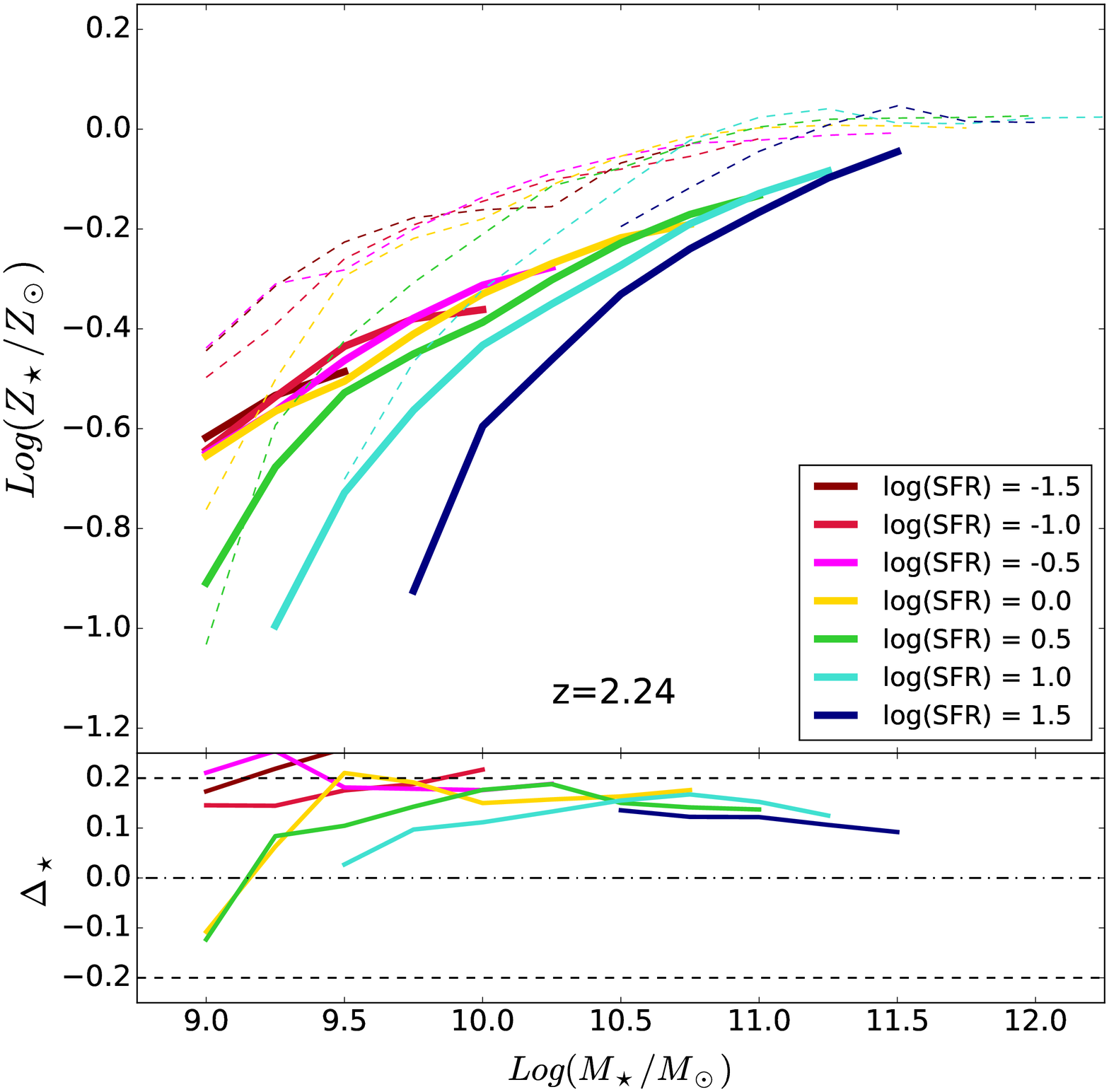} 
    \includegraphics[width=9cm]{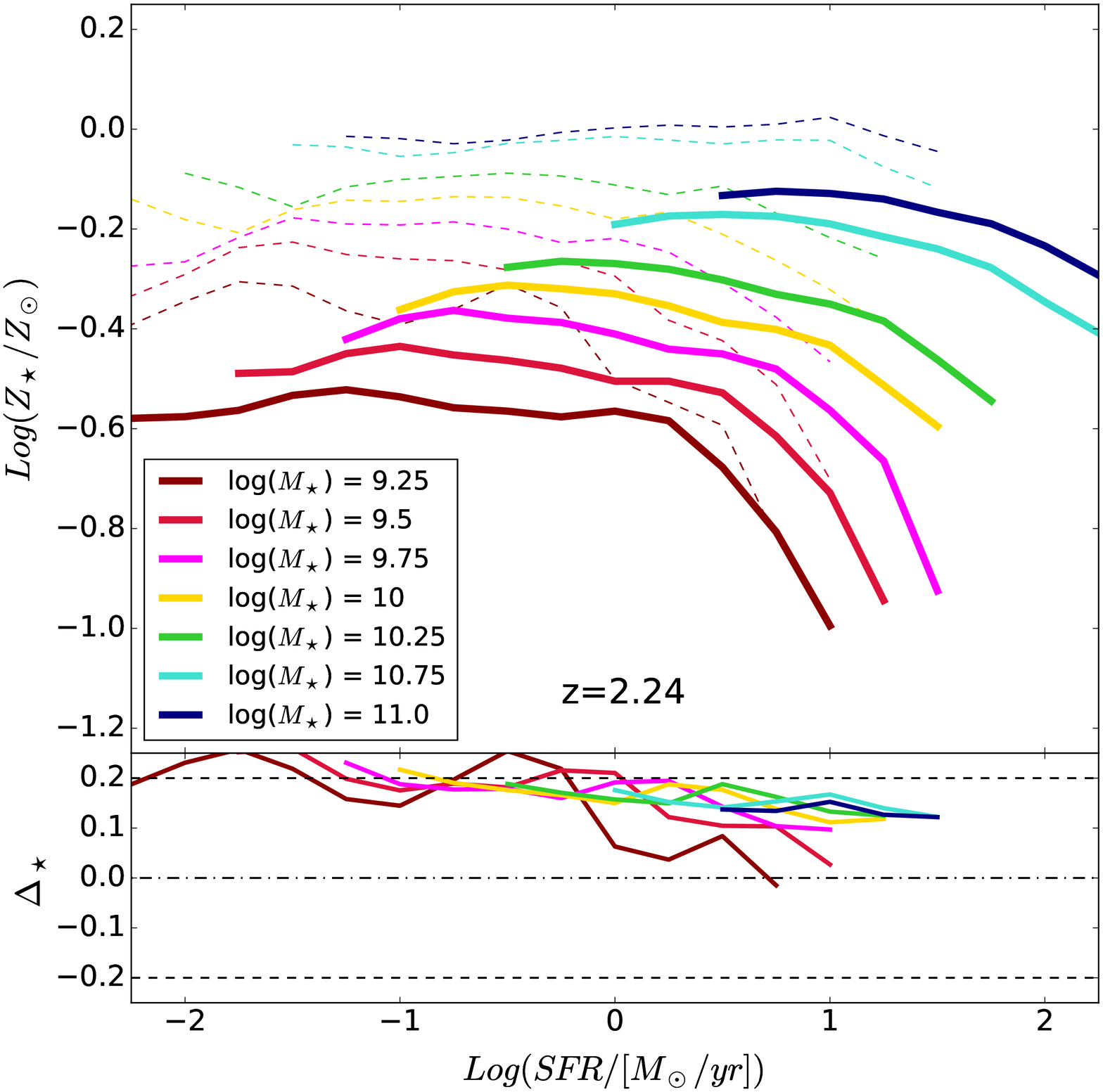} }
  \caption{Fundamental Metallicity Relation between SFR, $M_\star$ and
    cold gas metallicity at $z\sim2.25$ in {\gaea}. {\it Upper left
      panel:} the cold gas metallicity - stellar mass relation
    predicted in {\gaea} in different bins of $Log(SFR)$. {\it Upper
      right panel:} the cold gas metallicity - SFR relation predicted
    in {\gaea} in different bins of $Log(M_\star)$. {\it Lower left
      panel:} stellar mass-metallicity relation predicted in {\gaea}
    in different bins of $Log(SFR)$. {\it Lower right panel:} the
    $Z_\star$ - SFR relation predicted in {\gaea} in different bins of
    $Log(M_\star)$. In all four panels, thin dashed lines show the
    corresponding relations at $z=0$ and the lower insets report the
    differences $\Delta_g = Log(Z_{\rm gas}(z))-Log(Z_{\rm gas}(0))$
    and $\Delta_\star =
    Log(Z_{\star}(z))-Log(Z_{\star}(0))$.}\label{fig:manrel_224}
\end{figure*}

\section{Dataset}\label{sec:data}
\begin{figure}
  \centerline{ \includegraphics[width=9cm]{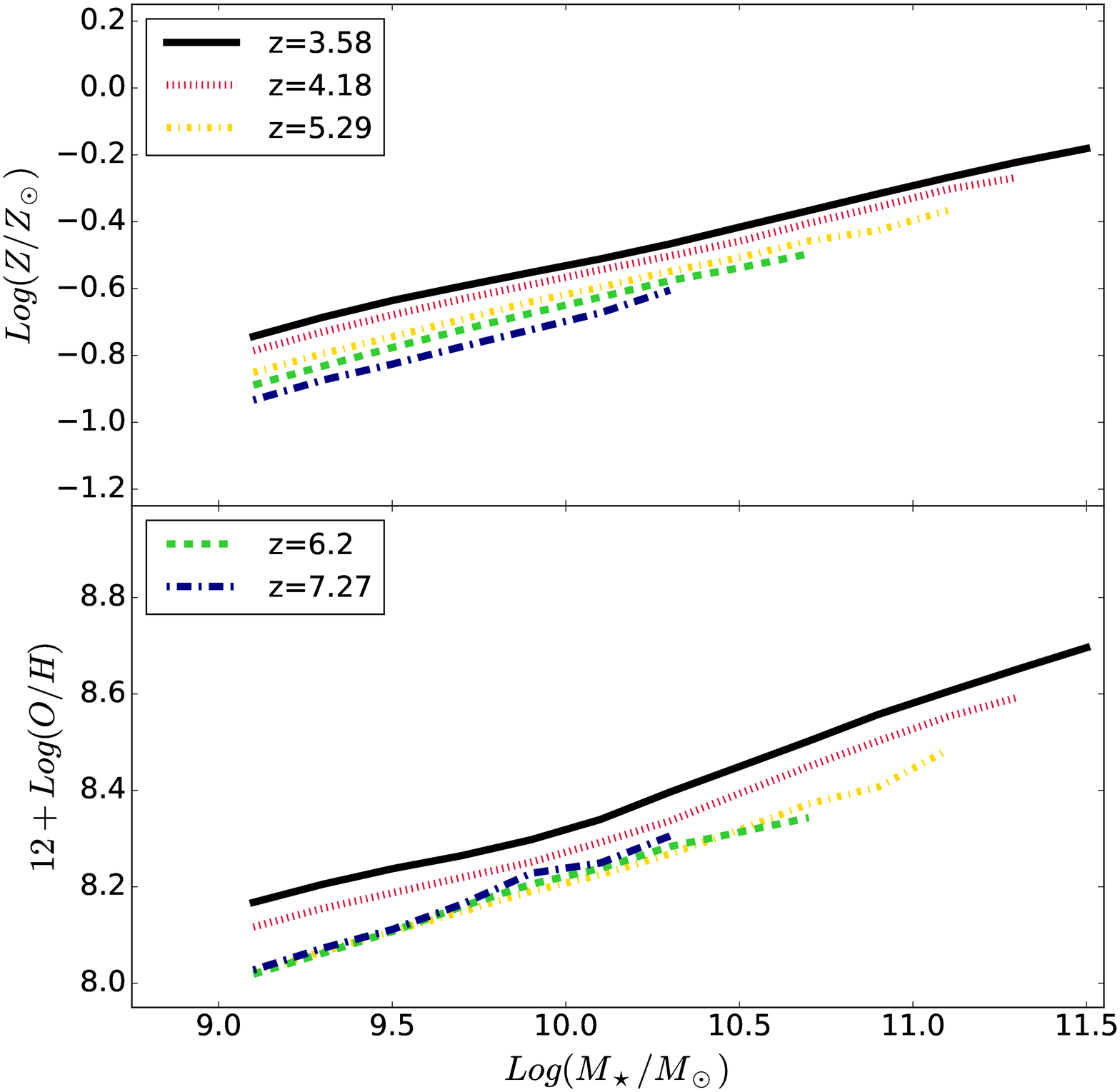} }
  \caption{Predicted evolution of the stellar (upper panel) and
    gas-phase (lower panel) MZR at $3.5 \lesssim z \lesssim
    7.2$.}\label{fig:jwst}
\end{figure}
We consider stellar metallicity measurements from the VANDELS ESO
public spectroscopic survey \citep{McLure18, Pentericci18} carried out
with the VIMOS spectrograph on ESO's Very Large Telescope (VLT). The
targeted fields are centred on the CANDELS CDFS and UDS, that already
include extensive WFC3/IR imaging \citep{Grogin11, Koekemoer11} and
for which photometric catalogues are available \citep[see
  e.g.][]{Guo13}. The main targets of the VANDELS survey are (a)
massive passive galaxies at $1<z<2.5$, (b) bright star-forming
galaxies at $2.4<z<5.5$ and (c) faint star-forming galaxies at
$3<z<7$. \citet{Pentericci18} provides more details on the
observations and reduction techniques. Stellar masses are derived from
SED fitting as described in \citet{McLure18}.

We take advantage of the results obtained by \citet{Cullen19} and
\citet{Calabro20} for the stellar MZR at $z\sim3.5$. \citet{Cullen19}
consider a sample of 681 star forming galaxies at $2.5 \lesssim z
\lesssim 5$ from the second VANDELS data release with robust redshift
determination. Since the typical S/N of VANDELS spectra is not high
enough to extract informations from individual sources, stacked
spectra in stellar mass and redshift are used. Metallicities are then
derived from full UV spectral fitting including indicators that trace
the stellar photospheric metallicity of young stellar
populations. Recently, \citet{Calabro20} propose an alternative
analysis for stellar metallicities in the most recent VANDELS sample
of 732 star-forming galaxies, that complements the \citet{Cullen19}
study. In particular, \citet{Calabro20} adopt different measurements
of stellar metallicity from individual UV absorption features. They
consider stacked spectra and restframe UV spectral features ($\lambda
1501$ and $\lambda 1719$). The two estimates are in good agreement: in
particular they share a similar dependence on stellar mass, increasing
$\sim$0.5 dex from $M_\star\sim10^9 \msun$ to $M_\star\sim10^{10.5}
\msun$. In order to put these results into a wider context, we also
consider results on the stellar phase metallicity coming from other
surveys at lower redshifts, such as \citet{Zahid17} and
\citet{Gallazzi14}. These studies are based on stellar metallicities
derived from optical spectral features, instead of the UV features
used in the VANDELS analysis. Optical features are sensitive to older
stellar populations than UV features. Furthermore, for consistency
with the VANDELS sample considered in this paper, we refer only to
results obtained from star-forming subsamples in \citet{Zahid17} and
\citet{Gallazzi14}.

In order to provide a more comprehensive picture of galaxy formation,
we consider not only the stellar MZR, but also the gas-phase MZR and
its redshift evolution. We thus complement VANDELS data with recent
results from the KMOS Lensed Emission Lines and VElocity Review
(KLEVER \citealt{Curti20z}). KLEVER employs the multi-object near-IR
integral field spectrograph KMOS on VLT. It focuses on the resolved
properties of 39 galaxies in the redshift range $1.2<z<2.5$. In
detail, the spectral resolution of the KMOS instrument allows an
accurate mapping of bright nebular emission lines at rest-frame,
including H$_\alpha$, H$_\beta$, [O{\sc ii}]$\lambda\lambda$3727,
3729\AA\AA, [O{\sc III}]$\lambda$ 5007\AA, [N{\sc II}] $\lambda$
6584\AA, [S{\sc II}] $\lambda\lambda$ 6717,31\AA\AA, [S{\sc III}]
$\lambda\lambda$ 9068,9530\AA\AA, fundamental for the estimate of the
metallicity of the cold gas. Stellar masses for the KLEVER sources
have been derived from SED fitting, while their SFRs are computed from
the estimated intrinsic H$_\alpha$ luminosity. In order to have an
homogeneous sample of gas-phase determinations, \citet{Cresci19}
collect gas-phase metallicities from several authors (see caption of
Fig.~\ref{fig:mzr} for all references) and convert all stellar masses
and SFRs determinations to the same Chabrier IMF. More importantly,
they re-evaluate individual gas-phase metallicities using the
strong-line (e.g. N$_{\rm II}$/H$_\alpha$) calibration as in
\citet{Curti20z} for all samples.

\section{Results}\label{sec:results}
\subsection{Mass-Metallicity relations}
We first consider the evolution of the mass-metallicity relations
(MZRs). In the left panel of Fig.~\ref{fig:mzr} we compare the
evolution of the stellar MZR in the redshift range $2.5 \lesssim z
\lesssim 5$ with data from VANDELS \citep{Cullen19, Calabro20,
  Talia20}, and with data at lower redshifts from the literature
\citep{Gallazzi14, Zahid17}. All data have been rescaled to a common
0.02 solar metallicity. {\gaea} predictions reproduce reasonably well
the local and the $z\sim0.7$ relations. The model predicts an
evolution of the MZR normalization to $z\sim3$ of roughly $\sim$0.35
dex in metallicity. This is roughly $\sim$0.25 dex less than the
evolution suggested by the~\citet{Cullen19} and~\citet{Calabro20}. We
stress however, that the tension between model predictions and the
highest-redshift data available is at 1-$\sigma$ level. It is
important to keep in mind that all observational results at this
redshift are based on the VANDELS dataset, but they differ both in
terms of sample size and analysis strategy. Nonetheless, all VANDELS
measurements rely on spectral features in the FUV spectral range. It
is expected that FUV-weighted metallicities could be offset low by
$\sim$0.1 dex with respect to mass-weighted metallicities \citep[see
  e.g.][]{Cullen19}. This is due to the fact that the youngest stellar
populations, that dominate the UV flux, are at the same time the most
metal rich and the most dust obscured. Moreover, FUV-weighted
metallicities mainly trace the iron abundance, as a proxy to total
metallicity, and this can also contribute to the offset from model
predictions. More interestingly, sources included in the
\citep{Cullen19} sample have a median sSFR of about $10^{-8.35}$
yr$^{-1}$, that is quite different from the typical
sSFR$\sim10^{-8.7}$ yr$^{-1}$ for z$\sim3$.5 galaxies in {\gaea}. We
will deepen this point in Section~\ref{sec:discussion}.  Overall, the
VANDELS stellar MZR shows a steeper slope than the stellar MZR
predicted by GAEA. \citet{Calabro20} study in detail the effect of
VANDELS selection criteria on the resulting slope of the $z\sim3.5$
MZR and show that when the same criteria are applied on {\gaea} mock
catalogues the predicted MZR slope is in better agreement with the
observed one (although the discrepancy in the normalization holds). As
in this paper we are mainly concerned about the predicted redshift
evolution of the global MZRs, we do not consider selection effects
while dealing with individual datasets.
\begin{figure}
  \centerline{ \includegraphics[width=9cm]{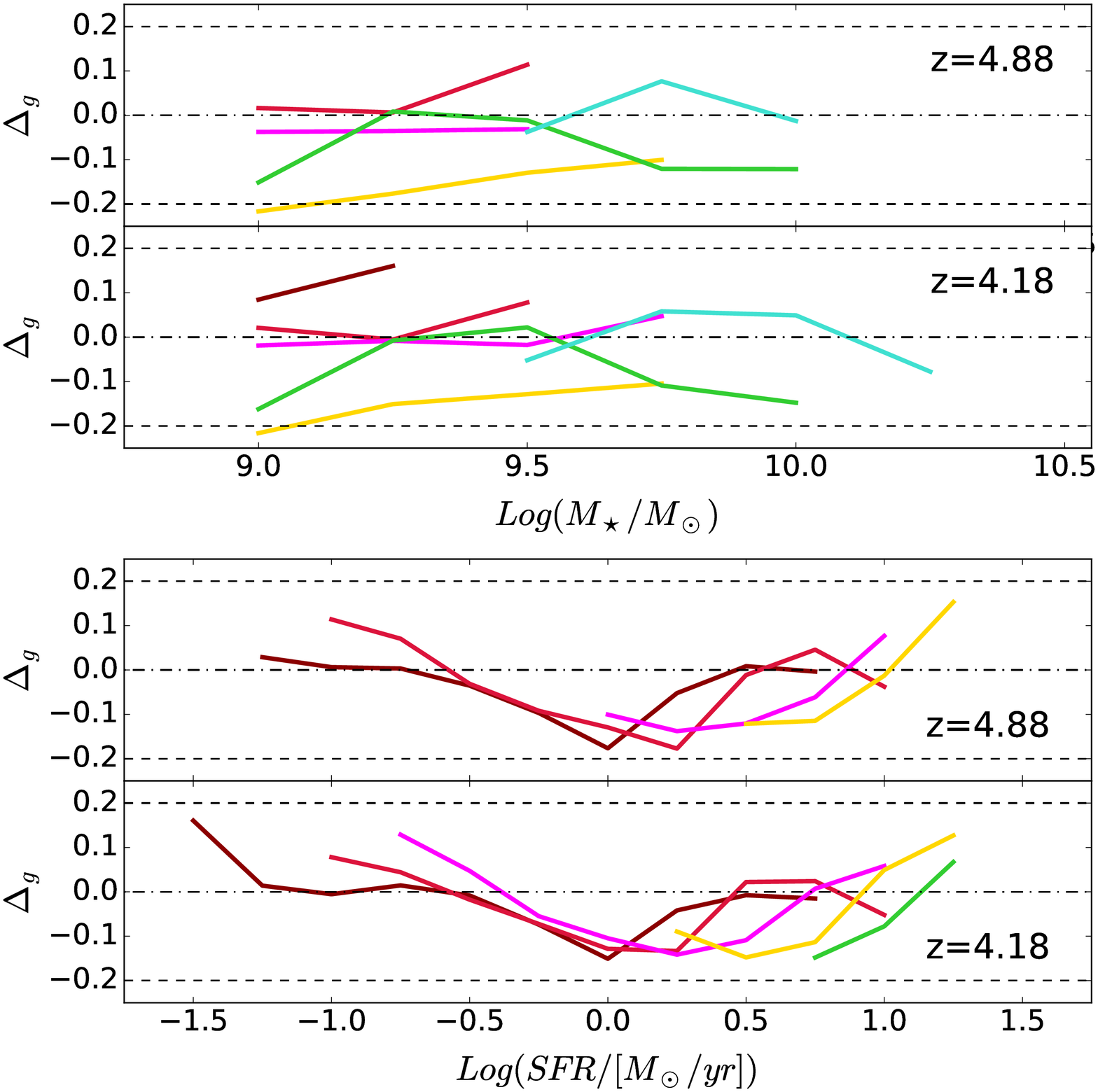} }
  \caption{Evolution of the gas-phase FMR in the redshift range
    $z\sim4-5$. The various panels show the evolution with respect to
    the $z=0$ FMR using the $\Delta_g = Log(Z_{\rm gas}(z))-Log(Z_{\rm
      gas}(0))$ indicator as in the lower insets in
    Fig.~\ref{fig:manrel_224}, and using the same colour
    scheme.}\label{fig:fmr_highz}
\end{figure}

We then consider the evolution of the gas metallicity as a function of
the stellar mass in the right panel of Fig.~\ref{fig:mzr}. We compare
{\gaea} predictions with a variety of data from several surveys as
indicated in the legend. We remind the reader that in this paper we
are considering the homogeneized values from \citet[][filled
  symbols]{Cresci19}, i.e. after reporting all individual datasets to
the same calibration framework used in \citet[][see previous section
  for more details]{Curti20z}. The advantage of such a large
compilation of homogeneized data lies in the possibility of studying
in detail the absolute evolution of the MZR, under the hyphothesis
that all data suffer from similar systematics at all redshift
\citep[see also][]{Maiolino08}. Moreover, the still relevant
calibration uncertainties between the several spectral indicators
adopted in the literature limit the constraining power of single epoch
MZRs to the shape of the relation. Therefore, the relative evolution
of the metal enrichment at fixed stellar mass at different cosmic
epochs represents a more fundamental test for models of galaxy
evolution. It is worth stressing that in \citet{Hirschmann16}, we
compared {\gaea} predictions with the \citet{Tremonti04} data for the
$z\sim0$ gas-phase MZR, that are systematically higher than
\citet{Curti20a}. In \citet{DeLucia20}, we compare {\gaea} predictions
with both \citet{Tremonti04} and \citet{Curti20a}
measurements. {\gaea} intrinsic predictions are indeed systematically
higher than the collection by \citet{Cresci19} (thin dashed lines in
Fig.~\ref{fig:mzr}, right panel). In order to focus on the redshift
evolution of the MZR it is therefore useful to shift downwards {\gaea}
predictions by 0.1 dex, based on the match to the $z\sim0$ data. It is
important to stress, that this shift does not correspond to any
retuning of the model parameters with respect to
\citet{Hirschmann16}. Fig.~\ref{fig:mzr} clearly shows that, under
this hypothesis, {\gaea} is able to recover the overall evolution of
the gas-phase MZR up the highest redshifts available and the predicted
scatter is also compatible with the available datasets, although some
tensions with the predicted slope of the relations hold (e.g. at z=0
it is somewhat steeper than observed, \citealt{DeLucia20}). It is
therefore important to stress that our results do not imply that a
constant shift of the predictions by the same amount is enough to
account for all the systematics in the observed data. In particular,
we do not expect these systematics to be the same at all galaxy mass
scales, but they should depend also on the mass scale under scrutiny
and affect the overall slope of the relation as well (as a matter of
fact the predicted slopes do not match perfectly the observed
determinations even after renormalization). Nonetheless, we deemed
that our approach allows us to explore if {\gaea} is able to predict
the same evolutionary trends in the observed data (irrespective of the
true overall normalization), both as a function of redshift and
stellar mass. Moreover, the 0.1 dex shift is acceptetable, given the
expected uncertainty in the normalization of the MZR, as due to the
calibration of the spectral indicators used to derive gas-phase
metallicities and to the set of reference theoretical templates
\citep{KewleyEllison08}. In particular, \citet{Curti20z} argue that
the electron temperature method could suffer from a systematic
underestimate of 0.1 dex, due to the presence of temperature
fluctuations in the HII regions. Therefore, we conclude that {\gaea}
is able to reasonably reproduce the observed evolution rate in both
stellar and gas-phase metallicity from $z~\sim3-3.5$ to $z\sim0$, that
is a more stringent constraint for galaxy evolution than the overall
normalization of the relation, that may also depend on the assumption
made for the metallicity estimate \citep{KewleyEllison08}.

While working on this paper, \citet{Sanders20} published new results
on the gas-phase MZR in the range $2.3<z<3.3$, obtained in the context
of the MOSDEF survey, using the Multi-Object Spectrometer For Infrared
Exploration (MOSFIRE) instrument. These metallicity determinations are
based on the analysis of spectral features on stacked spectra; however
it has not been possible to repeat the homogeneization procedure
discussed previously. For this reason, we show these results as open
circles in Fig.~\ref{fig:mzr}: it is worth noting that the MZR relation
they trace agrees quite well with the {\gaea} intrinsic relation
(dashed lines). This confirms both the intrinsic dispersion of the
MZRs based on different calibrations, and the fact that {\gaea} is
able to reproduce the redshift evolution of the MZR, as constrained by
homoegeneous MZR determinations.

\subsection{Fundamental Metallicity Relation}
\begin{figure*}
  \centerline{
    \includegraphics[width=9cm]{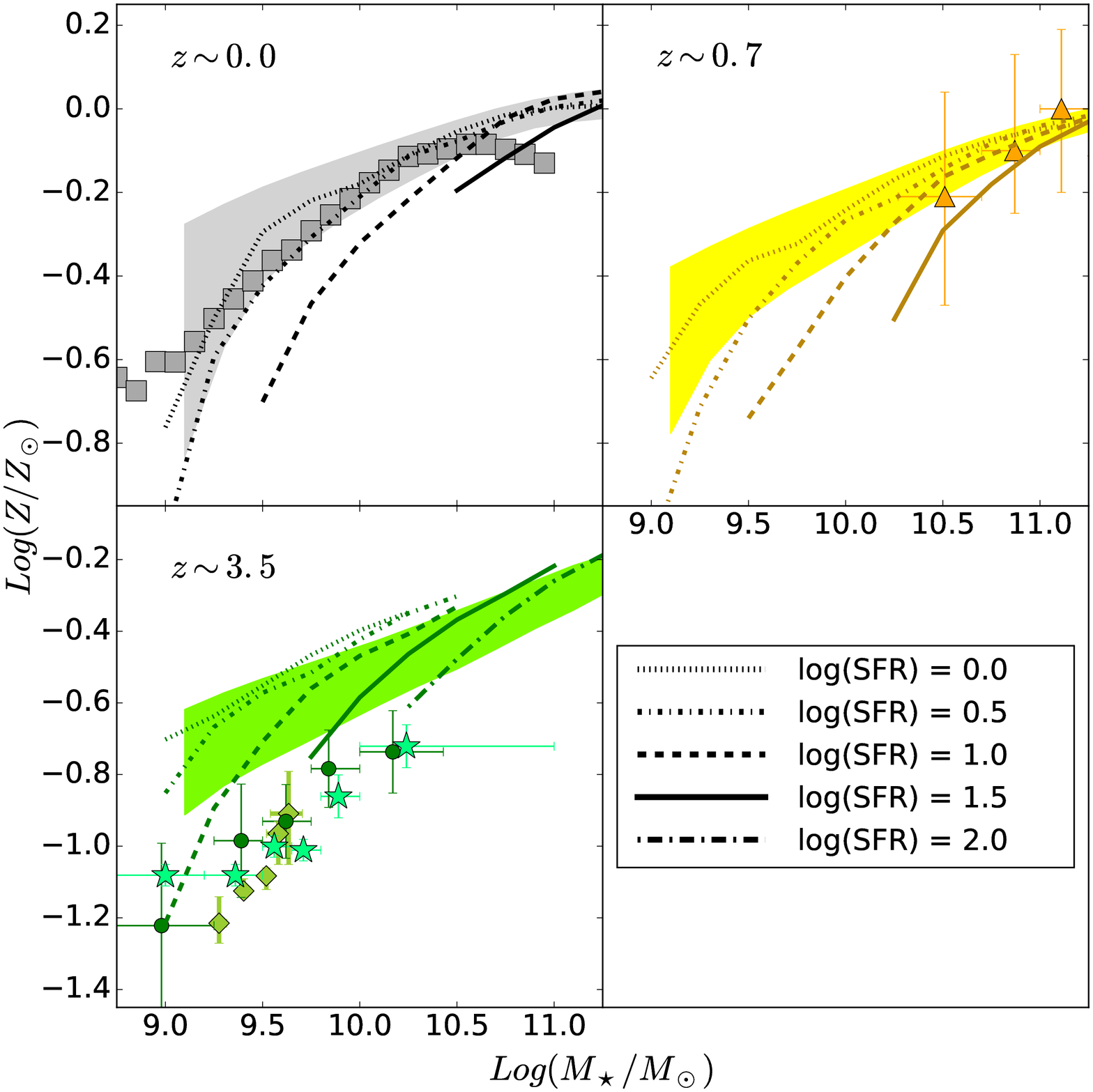}
    \includegraphics[width=9cm]{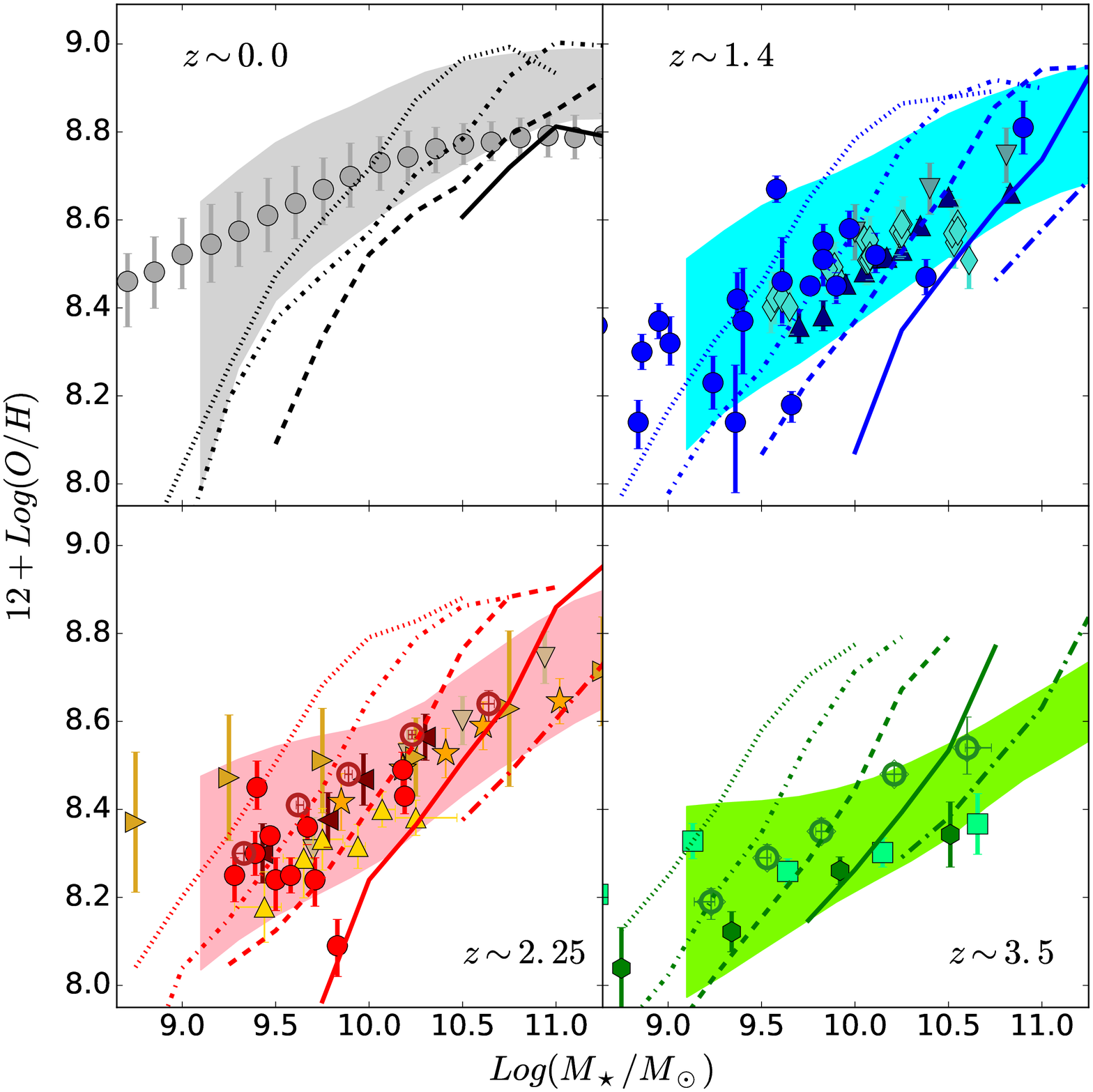} }
  \caption{Stellar (left panels) and gas-phase (right panels) MZRs for
    samples of galaxies at a given SFR level (as indicated in the
    legend). All other datapoints and shaded areas as in
    Fig.~\ref{fig:mzr}.}\label{fig:mzr_split}
\end{figure*}
\begin{figure*}
  \centerline{
    \includegraphics[width=9cm]{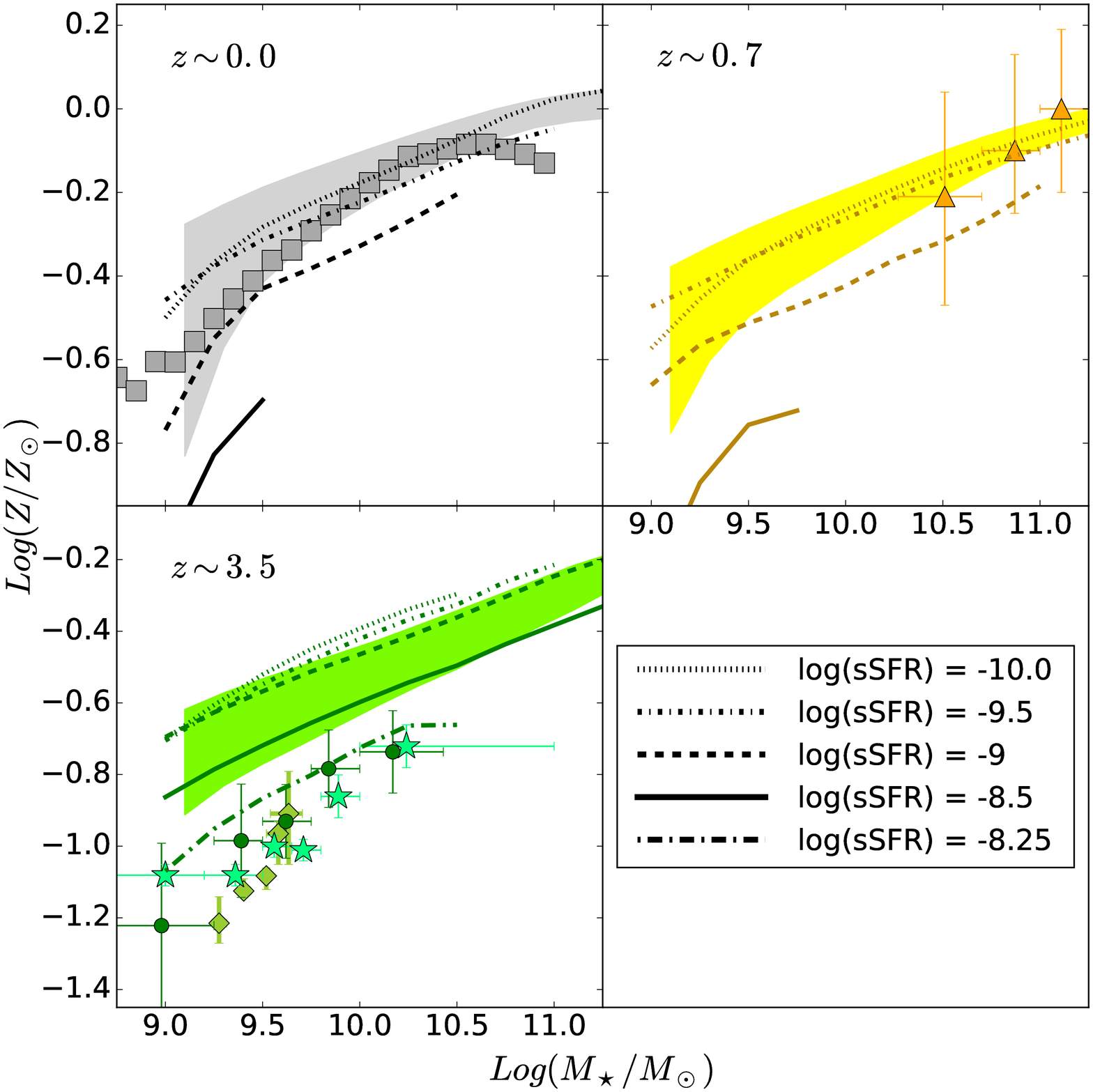}
    \includegraphics[width=9cm]{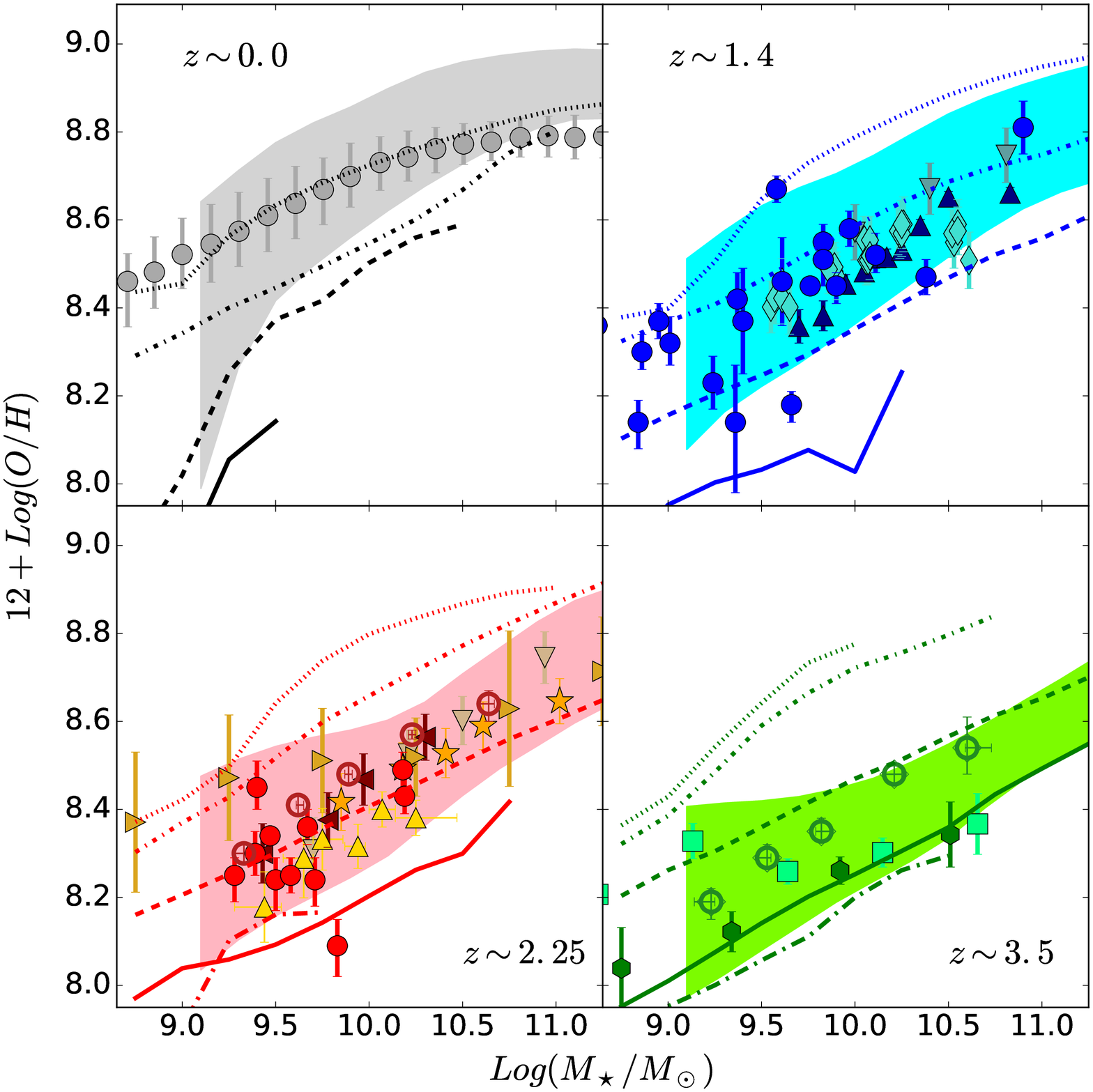} }
  \caption{Stellar (left panels) and gas-phase (right panels) MZRs for
    samples of galaxies at a given sSFR level (as indicated in the
    legend). All other datapoints and shaded areas as in
    Fig.~\ref{fig:mzr}.}\label{fig:mzrgas_split}
\end{figure*}
\begin{figure*}
  \centerline{
    \includegraphics[width=9cm]{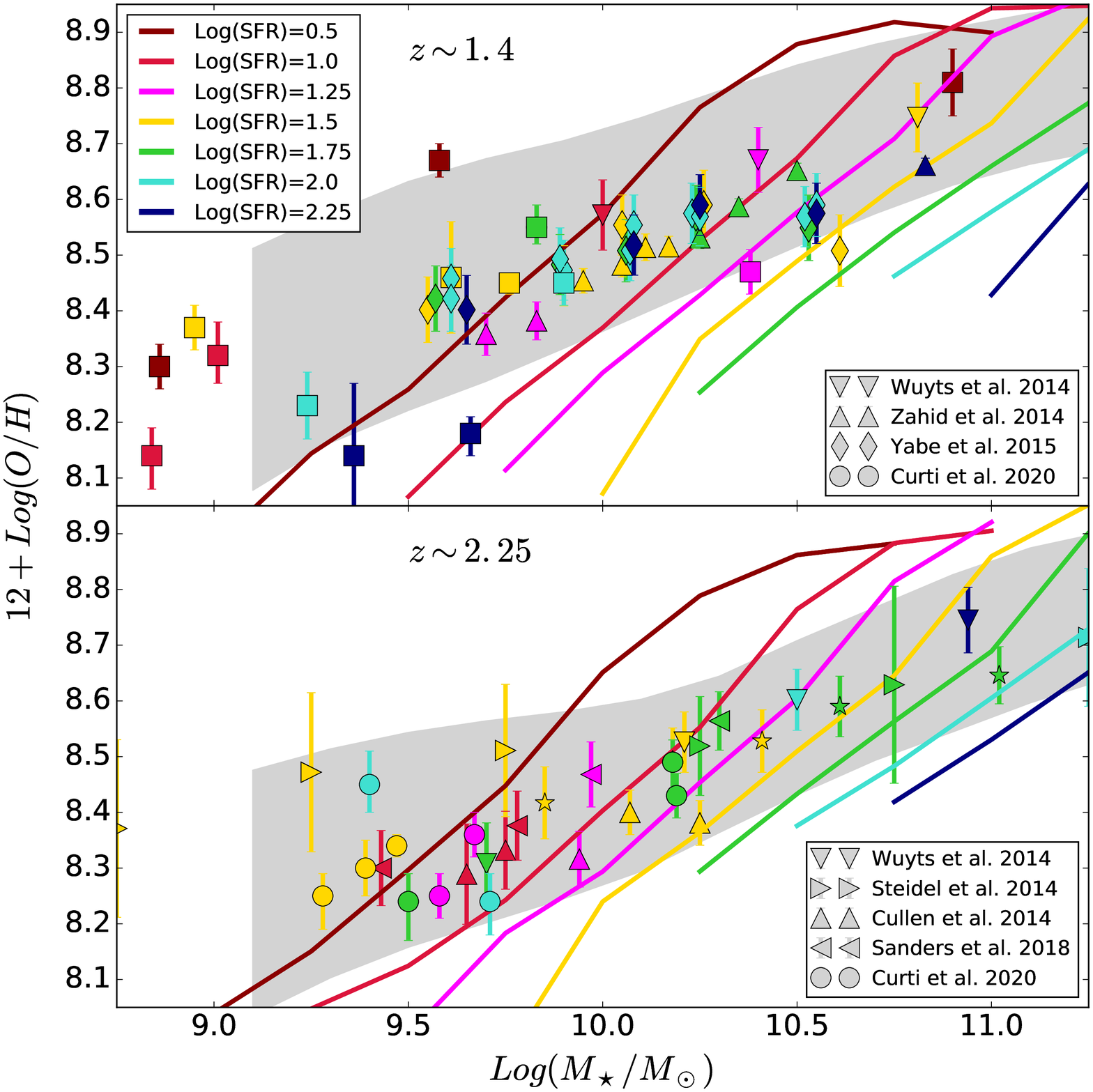} 
    \includegraphics[width=9cm]{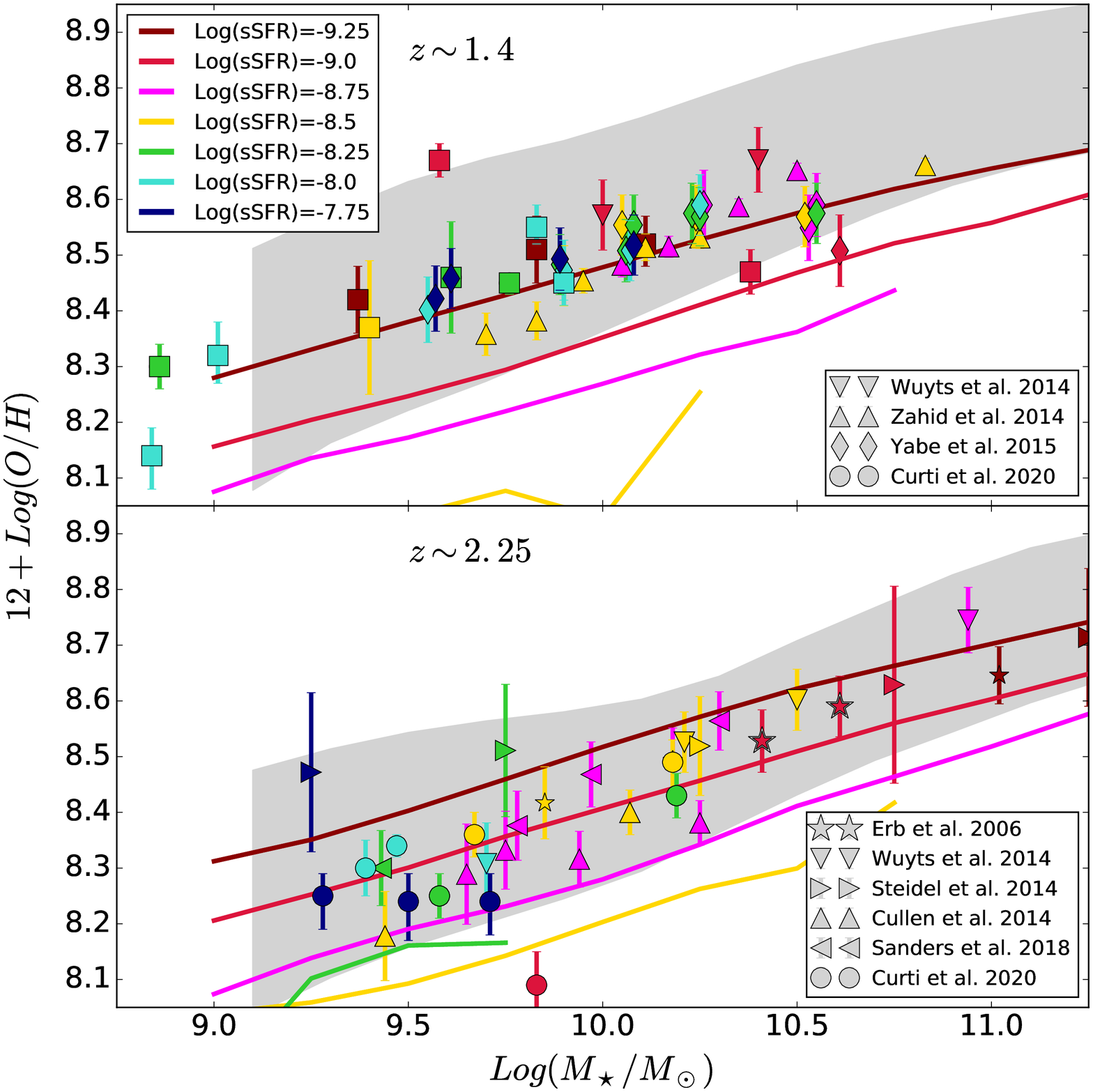} }
  \caption{Gas Phase MZR at $1.4<z<2.25$. In the left (right) panel,
    both data and model predictions have been split in colour
    according to their SFR (sSFR) levels, as indicated in the
    legend. The grey shaded area shows the predicted MZR as in
    Fig.~\ref{fig:mzr}. Only data with SFR (sSFR) inside the
    considered model range are shown.}\label{fig:sfr_split}
\end{figure*}
We then consider in detail the relation between the cold gas
metallicity, the SFR and $M_\star$. \citet{Mannucci10} first showed
that the mass-metallicity relations we discussed in previous
paragraphs are just a projection of a more general relation between
$M_\star$, SFR and cold gas metallicity (Fundamental Metallicity
Relation, FMR). Moreover, \citet{Mannucci10} also show that the FMR is
remarkably similar up to z=2.5, while the limited amount of data
available at higher redshifts points towards a significant evolution
\citep{Mannucci09, Troncoso14}. Indeed, almost all $z\lesssim2-3$
galaxies with accurate measurements of these properties lie within
0.6-0.2 dex from the $z\sim0$ FMR. Therefore, the observed evolution
in the mass-metallicity relations can be explained simply by assuming
that galaxies at different redshift sample different regions of the
FMR.

A detailed comparison between {\gaea} predictions and the $z=0$ FMR
(including \citealt{Curti20a}) has been presented in
\citet{DeLucia20}\footnote{It is worth noting that in
  \citet{DeLucia20} we use a model version based on predictions from
  the more recent version of the model from \citet{Xie17}, featuring
  prescriptions for splitting the cold gas phase into its HI and
  H$_2$ components derived from \citet{BlitzRosolowsky06}. The
  predictions presented here, on the other hand, derive from the model
  implementing the ``FIRE'' feedback prescription presented in
  \citep{Hirschmann16}: this represents our standard realization,
  which we show is able to provide a good agreement with the evolution
  of the GSMF and cosmic SFR up to $z\sim7$ \citep{Fontanot17b}. We
  stress that the main trends we discuss in this paper do not depend
  on the specific version of the model adopted.}: overall, the model
predicts a well defined FMR in reasonable agreement with
observations. There are, however, some evident discrepancies between
predictions and data: the most notable is the different metallicity
normalization (as we already mention in the previous
section). Moreover, the predicted slope of the $M_\star$-$Z_{\rm gas}$
relation at fixed $SFR$ is much steeper than observed for $SFR$ lower
than a few $\msunyr$. The predicted slope at the highest SFRs is in
better agreement with the observational constraints.

In this paper we focus on the redshift evolution of the FMR as
predicted by {\gaea}. In the upper panels of Fig.~\ref{fig:manrel_224}
we show the FMR predicted by {\gaea} at $z\sim2.24$ (solid lines). We
chose this redshift as a reference since this is representative of the
shape of the relation in the $1.5<z<3.5$ range. The left panel shows
the $Z_{\rm gas}$-$M_\star$ relations at fixed SFR, while the right
panel shows the $Z_{\rm gas}$-SFR relations at fixed $M_\star$. We
then compare the shape of the distributions with the corresponding
predictions at $z=0$ (thin dashed lines in all panels).

In detail we consider in the lower insert of each panel the residuals
with respect to the $z=0$ FMR:

\begin{equation}
\Delta_{\rm g} = Log(Z_{\rm gas}(z))-Log(Z_{\rm gas}(0)).
\end{equation}

\noindent
where $Z_{\rm gas}(z)$ and $Z_{\rm gas}(0)$ represent the {\gaea}
predicted MZRs at the considered redshift and at $z=0$. For both
projections the deviation from the $z=0$ relations is smaller than
0.1-0.2 dex, in agreement with the observational evidence from
\citet{Mannucci10} and~\citet{Curti20z} for high redshift galaxies. We
confirm that model galaxies populate a well defined FMR already at
intermediate redshifts, and, in particular, the $z=0$ relations is
already in place at $z\sim2$.

We then consider another possible FMR, defined by using the stellar
metallicity (lower panels of Fig.~\ref{fig:manrel_224}). A {\it
  stellar} FMR is well defined at all redshift in {\gaea}, with a
shape similar to the gas-phase FMR. However, while the gas-phase FMR
is characterized by a defined shape {\it across all redshifts} in the
$M_\star$-$Z_\star$-SFR parameter space, the stellar FMR is
characterized by a marked evolution of its overall normalization with
redshift. Nonetheless, in the following we will refer to a ``stellar
FMR'' in order to highlight its analogy with the original gas-phase
FMR, but keeping in mind the different nature of this
redshift-dependent relation. In particular, from $z\sim2.25$ to $z=0$,
the shift is already of the order of $\sim$0.2 dex in metallicity (we
define $\Delta_\star$ analogously to $\Delta_g$), and keeps increasing
at higher redshift (reaching $\sim$0.3-0.4 at z$\sim$5). These results
thus suggest that {\gaea} predicts a universal gas-phase FMR.  On the
other hand, the stellar FMR traces the increasing baryonic mass locked
in the metal component as galaxies evolve.

{\gaea} allows us to provide predictions for the FMR at even higher
redshifts, that will be possible to test thanks to future facilities
like the James Webb Space Telescope (JWST). In particular, JWST is the
only instrument that will be able to access the strong lines needed
for measuring gas-phase metallicities beyond
$z\sim3.5$. Fig.~\ref{fig:jwst} shows the expected evolution of the
total stellar and gas-phase MZR at $3.5 \lesssim z \lesssim 7.2$. The
stellar MZR is characterized by a steady evolution of the
normalization and a constant slope, while the gas-phase MZR seems to
evolve only up to $z\sim5$. For both relations, the degree of
predicted evolution is relatively small ($\sim$0.2 dex), and of the
same order of magnitude as the intrinsic scatter in observed samples
at lower redshifts. Our model thus suggests that it will extremely
difficult to constrain any evolutionary trend at these redshifts, even
using JWST observations. Finally, in Fig.~\ref{fig:fmr_highz}, we show
the evolution of the gas-phase FMR in the redshift range
$z\sim4-5$. All $\Delta_g$ are still within 0.2 dex from the $z=0$
FMR, thus showing that this scaling relation is already well defined
in {\gaea} up to the highest redshift.

\section{Discussion}\label{sec:discussion}
In previous sections, we discussed the MZRs and the FMR (and their
redshift evolution). In Fig.~\ref{fig:mzr_split} we consider the same
data as in Fig.~\ref{fig:mzr}, but we show the MZR at fixed SFR (as
marked in the legend). The MZR derives from the linear convolution of
these individual relations, weighted by the space densities of
galaxies with the corresponding SFR. The difference between the
individual lines shown in Fig.~\ref{fig:mzr_split} and the global
relations show in Fig.~\ref{fig:mzr} (especially at high-redshifts)
reflects the redshift evolution of the relative contribution of
galaxies at fixed SFR. Fig.~\ref{fig:mzr_split} clearly shows that
{\gaea} predicts a MZR at fixed SFR much steeper than the global MZR,
for both the stellar and gaseous components. This is particularly
evident for the gas-phase, where none of the individual MZR at fixed
SFR has a slope similar to that of the global MZR. Observational
measurements of the FMR \citep[see e.g.][their Fig.~6]{Curti20a}
indeed favour steeper slopes for high-SFR galaxies; the slope then
shallows, becoming compatible with the slope of the global MZR below 1
$\msunyr$. The invariance of the MZR slope in {\gaea} at fixed SFR
seems to suggest that the balance between gas cooling, star formation
and feedback processes, as modeled by the SAM, works at all stellar
mass scales \citep[see e.g.][]{FinlatorDave08}. The observed slope
evolution may thus hint that either a different equilibrium
configuration is in place at different mass scales or that additional
physical processes are at play for low-mass galaxies. Another possible
caveat is related to the fact that in Fig.~\ref{fig:mzr_split}
and~\ref{fig:sfr_split} we are showing the intrinsic {\gaea}
predictions: convolving these with an estimate of the typical error in
the SFR and $M_\star$ determinations has the effect of flattening out
the intrinsic relations \citep{DeLucia20}, thus easing the discrepancy
with the data.

In Fig.~\ref{fig:mzrgas_split}, we consider the MZRs at fixed
sSFR. When we split according to the sSFR, the MZRs are characterized
by slopes more similar to the global MZR, both in the stellar and
gas-phases. In particular, these plots clearly indicate the typical
sSFR of the galaxy population dominating the MZRs. The slope of the
individual MZR at fixed sSFR are pretty similar to each other, in
reasonable agreement with the observational trends, that show only a
small evolution \citep{Curti20a} with sSFR. The normalization of the
MZR shows a decreasing trend at increasing sSFR, which brings the
sSFR$>10^{-8.5}$ yr$^{-1}$ relations in better agreement with the
z$\sim$3.5 VANDELS measurements. It is thus tempting to explain the
discrepancy between the VANDELS and {\gaea} MZR as an effect of the
different typical sSFR of the underlying population. However,
sSFR$>10^{-8.5}$ yr$^{-1}$ galaxies in {\gaea} represents a population
well separated from the SF main sequence, as can be appreciated by
comparing the MZR at fixed sSFR with the shaded area representing the
range between 16th and 84th precentiles of the distribution. We do not
expect VANDELS galaxies to be that systematically offset from the
z$\sim$3.5 main sequence \citep{Garilli21}.

In Fig.~\ref{fig:mzr_split} and~\ref{fig:mzrgas_split}, we consider
SFR and sSFR ranges wide enough to include all galaxies in our model
sample. However, data at intermediate redshift ($1.4<z<2.5$) allow for
a direct comparison in a comparable SFR range. In the left panel of
Fig.~\ref{fig:sfr_split} we colour-coded the MZR at fixed SFR and we
then mark with similar colour the position of individual galaxies
(only data with SFR or sSFR inside the considered model range are
shown). Whereas it is difficult to find a clear FMR trend from such a
small galaxy sample, it is clear that {\gaea} predictions tend to
underpredict galaxy metallicity at fixed SFR, especially at $M_\star
\sim 10^{10} \msun$. Furthermore, there is a clear difference in the
stellar mass range covered at fixed SFR in model predictions and
observations. While {\gaea} predicts a clear correlation between
$M_\star$ and SFR (with the largest SFR available only for more
massive galaxies), the range of observed $M_\star$ at fixed SFR is
typically much larger and irrespective of the actual SFR value. This
discrepancy may be connected to the different way of estimating SFR in
{\gaea} and in the data. However, it is worth noting that the
observational estimates are mainly based on H$_\alpha$ lines that
sample recent SFR ($<10^7$ yrs) that are comparable with the model
SFRs, which are computed over the time-span between two snapshots in
the simulations (roughly covering a few $10^8$ yrs). When considering
the right panel of Fig.~\ref{fig:sfr_split}, the data do not show any
clear trend in sSFR as {\gaea} predictions clearly do. In particular
the model has a hard times reproducing the metallicity of the highest
sSFR ($>10^{-8.5} yr^{-1}$) in moderate mass $< 10^{10} \msun$
galaxies. Of course, also in this case the different SFR estimate
between data and models should play a role in the comparison.

\begin{figure}
  \centerline{ 
    \includegraphics[width=9cm]{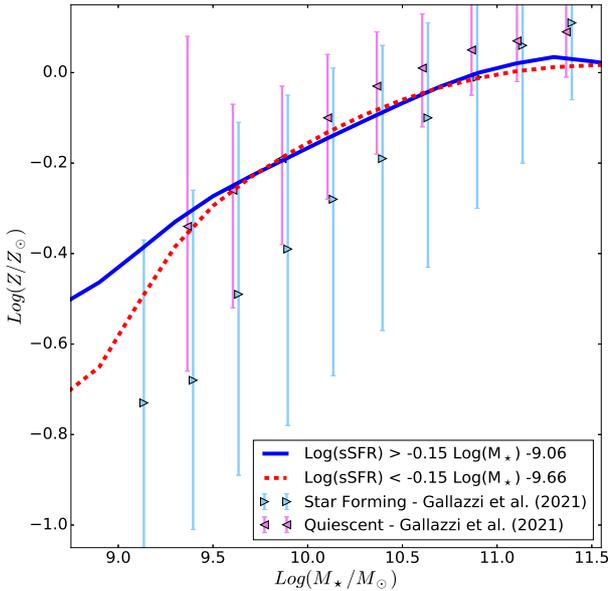} }
  \caption{$z=0$ stellar MZR for star-forming and quiescent
    galaxies. Both SDSS data (triangles with errorbars) and model
    galaxies (colored lines) have been separated into star-forming and
    quiescent samples using the sSFR cuts defined in
    \citet{Gallazzi20} and indicated in the legend. Errobars represent
    the 16th and 84th percentiles of the
    distribution.}\label{fig:peng}
\end{figure}

As a final check, we also consider the stellar MZR for quiescent and
star-forming galaxies \citep{Peng15, Trussler20}. In particular, in
Fig.~\ref{fig:peng} we consider recent results using the SDSS sample
from \citet{Gallazzi20}. They confirm a clear trend of lower stellar
metallicity at $M_\star < 10^{11} \msun$ for star-forming galaxies
with respect to quiescent sources. We then compare these data against
{\gaea} predictions using the same sSFR cut as in \citet[see figure
  legend for more details]{Gallazzi20}. At variance with SDSS data,
{\gaea} predicts an almost indistinguishable MZR for quiescent and
star-forming galaxies. We can understand these results in the light of
our previous analysis: {\gaea} predicts no slope variation for the MZR
at fixed sSFR, and the assumed activity threshold level at $z\sim0$ is
too low for the normalization between the resulting populations to be
different. We thus interpret this result as the effect of the tension
between the observed and predicted FMR, and in particular to the fact
that the slope of the FMR at fixed SFR does not evolve in {\gaea} in
the $M_\star < 10^{11} \msun$ mass range.

\subsection{Comparison with other theoretical models.}

In this section, we will put the results we discuss in this paper into
the context of other recent theoretical models. The main problem in
such a comparison lies in the fact that the different groups use
different observational samples as benchmark to assess the reliability
of their model MZR predictions. Given the uncertainties on the
absolute normalization of MZR (relative to the different spectral
features used), this translates into a limited quantitative comparison
between the models.

Within the SAM framework, \citet{Yates12} discuss the evolution of the
MZRs in the context of the \citet{Guo11} model: they find that the
$z=0$ gas-phase MZR is well reproduced by model predictions, but it
shows negligible evolution to higher redshifts. This is due to a rapid
enrichment of the cold gas in the model, that brings galaxies onto the
local relation already at $z\sim3$. This behaviour is connected with
the too efficient formation of $M_\star<10^{11} \msun$ galaxies in
previous generations of SAMs and hydro-simulations (see
e.g. \citealt{Fontanot09b, Weinmann12}). A more recent rendition of
the model \citep{Yates21} shows a more relevant evolution of the MZR
of the order to 0.3 dex from $z\sim3$ to $z\sim0$, as an effect of the
assumed highly efficient metal removal from galaxies via outflows.

The evolution of the MZRs and their relation with the FMR is the focus
of recent work in the framework of hydro-simulations. The {\sc EAGLE}
simulation \citep{Schaye15} is able to reproduce the local gas-phase
MZR and FMR. \citet{Lagos16} and \citet{DeRossi17} further expand the
analysis to higher-redshifts, by focusing in particular on the
redshift evolution of the gas-phase MZR. They find that the evolution
of the gas-phase MZR is well reproduced up to $z\sim1.5$. At higher
redshifts the situation is less clear, and they show that {\sc EAGLE}
predicts too large metallicities at $z>3$ with respect to the
observational estimates from \citet[][and based to a subsample of the
  \citealt{Troncoso14} data used in this paper]{Maiolino08}. However,
they also point out that the predicted evolution is in better
agreement with the observed evolution up to $z\sim3$ in the
Metallicity Evolution and Galaxy Assembly (MEGA) dataset
\citep{Hunt16}. Interestingly, they also find the FMR to be already
established at $z\sim5$, in agreement with our results.

\citet{Torrey19} study the evolution of the gas-phase MZR in the
context of the {\sc IllustrisTNG} simulation \citep{Pillepich18}, and
find that the model reproduces its slope and normalization evolution
up to $z\lesssim2$. At higher redshifts, the same considerations and
caveats we discussed for {\sc EAGLE} apply. An important aspect in the
{\sc IllustrisTNG} framework lies in the fact that the model assumes
that stellar-driven galactic winds are metal depleted with respect to
their parent ISM. This choice originates from the assumption that they
are hydro-dynamically decoupled for some time, in order to account for
the entrainment of wind material with the lower-metallicity
circum-galactic medium. At variance with this approach, stellar driven
winds in {\gaea} retain the same metallicity as the ISM of the parent
galaxy.

{\sc SIMBA} \citep{Dave19} is a cosmological hydro-dynamical
simulation that implements a model for stellar feedback derived from
the fitting formulae proposed in \citet{Muratov15}. In
\citet{Cullen19}, we show that this realization underpredicts the
stellar metallicity levels found in VANDELS $z\sim3.5$ galaxies,
although it correctly reproduces the shape of the observed MZR
relation. Unfortunately, in that paper {\sc simba} was not run down to
$z=0$, and no other datasets was available in the redshift range
covered by the simulation, so that we cannot explicitly check the
predicted evolution of the stellar MZR.

\citet{Cullen19} also present a comparison of their stellar MZR
with the corresponding relation defined by the {\sc FIRE} suite of
high-resolution zoom simulations \citep{Ma16}, showing that these
model galaxies correctly reproduce the amount of evolution from
$z\sim3-4$ to $z\sim0$. It is important to keep in mind that the
\citep{Ma16} sample is based on simulations of isolated dark matter
haloes, therefore it lacks the statistical information on cosmological
volumes. Nonetheless, the overall normalization of the theoretical
MZRs has to be shifted upwards by 0.3 dex to match the actual
observations, highlighting once more the difficulties of such models
to reproduce the correct level of chemical enrichment in individual
galaxies. \citet{Cullen19} conclude that shape and normalization of
the MZR in hydro-simulations are determined by the strength of
galactic outflows: models consistent with the FIRE feedback scaling
relations typically underpredicts the normalization of the MZR, while
earlier schemes with weaker galactic wind (like for example
\citealt{DallaVecchiaShaye12}) tend to overpredict galaxy
metallicities at higher redshifts. Our results based on {\gaea}
confirm on a cosmological volume that models implementing scaling
relations for stellar feedback based on the FIRE simulations are
indeed able to provide a reliable prediction for the evolution of the
MZR relations (both in stellar and gas-phase). Moreover, we show that
our prescriptions calibrated on reproducing the GSMF are also able to
recover the normalization of the MZR to within $\sim$0.1 dex.

\section{Summary}\label{sec:final}

In this paper, we present the comparison of the predictions of the
semi-analytic model {\gaea} with the $z\sim3.5$ stellar MZR
measured in the VANDELS framework. In order to put these results in
the broader context of galaxy formation and evolution we also consider
other determinations of the stellar MZR at lower redshift and the
redshift evolution of gas-phase MZR at $z<3.5$. Finally, we also
consider the evolution of the FMR defined by the 3-dimensional
$M_\star$, SFR and metallicity space.

We show that {\gaea} is able to reproduce the evolution of the
gas-phase MZR from $z\sim3.5$ to present, although some recalibration
of model predictions might be required to match the absolute scale
without resorting to a constant shift in the overall
normalization. However, this small shift is perfectly in line with the
uncertainty in the observational determination of metallicities. The
agreement between model predictions and the observed stellar MZR is
good up to $z\sim0.7$, while some tensions (at 2-$\sigma$ level) with
the results of the VANDELS sample are evident at $z\sim3.5$, in terms
of both the slope and normalization of the stellar MZR. A likely
solution for the slope tension lies in the selection criteria adopted
in VANDELS. If VANDELS target have been selected with a SFR higher
than the typical value in {\gaea}, that may explain the different
slope for the MZR \citep{Calabro20}. The different normalization of
the MZR can be partly explained by the fact that $Z_\star$ estimates
in VANDELS are FUV-weighted and this may lead to an offset to lower
metallicities with respect to mass-weighted metallicities
\citep{Cullen19}. A more quantitative comparison between model
predictions and observational determination requires to expand {\gaea}
in order to predict realistic synthetic spectra including all the
relevant spectral features for metallicity determination in the
different phases \citep[see e.g.][]{Hirschmann17, Hirschmann19}. These
synthetic features could be than processed in the same way as the real
data, and provide fundamental clues on the relation between intrinsic
and measured metallicities. We plan to explore these ideas in
forthcoming works.

Furthermore, {\gaea} predicts a steeper slope for the MZR at fixed
SFR, i.e. the FMR. As shown in \citet{DeLucia20}, these predictions
are in reasonable agreement with the observational determinations
\citep{Curti20a} for high-SFR galaxies, while some tension arise for
sources with lower-levels of activity: below a few $\msunyr$ data
suggest a flattening of the slope of the MZR at fixed SFR, while
{\gaea} predicts a constant slope (and steeper than the overall
MZR). This is probably the strongest discrepancy between model
predictions and the available observational constraints. However, it
is also worth noting that a possible way to reconcile data and models
lies in considering the effect of the uncertainties in the SFR
determination, when estimating the slope of the relation. Larger SFR
errors may lead to flatter slopes than the intrinsic value, so that an
increasing error at increasing SFR would mimic a slope evolution.

Furthermore, we use {\gaea} realizations to explore the redshift
evolution of the FMR. {\gaea} predicts the gas-phase FMR to be already
well established at $z\sim5$: this implies that the observed evolution
of the gas-phase MZR are mainly driven by typical galaxies at a given
redshift populating different regions of the FMR. On the other hand,
the stellar FMR shows a marked evolution in its normalization, with
galaxies increasing their $Z_\star$ content with cosmic time. These
predictions will be challenged with future facilities like JWST, which
will be able to open the $z\gtrsim4$ window to metallicity studies
thanks to spectral range covered by the NIRSPEC instrument. In the
meantime, stellar metallicity studies using some of the most recent
Multi-Object Spectrograph facilities (such as WEAVE, 4MOST, MOONS,
MOSAIC) or surveys like LEGA-C \citep{vanderWel16} will be able to
close the gap at intermediate to high redshifts and give us a better
understanding of the MZR evolution at $1\lesssim z \lesssim3$, also
taking into account the known uncertainties in the derivation of
stellar metallicities from spectral analysis.

These results confirm {\gaea} is able to reproduce, {\it at the same
  time}, the evolution of the GSMF up to $z\sim7$, the cosmic SFR up
to $z\sim10$ and the evolution of the stellar content up to
$z\sim3.5$. This is an important confirmation for our approach to
stellar feedback, highlighting once more the relevance of
stellar-driven winds in shaping galaxy evolution.
 
\section*{Acknowledgements}
This work is based on data products from observations made with ESO
Telescopes at La Silla Paranal Observatory under ESO programme ID
194.A-2003 (PIs: Laura Pentericci and Ross McLure) and ESO Large
Program 197.A-0717 (PI: Michele Cirasuolo). RA acknowledges support
from ANID FONDECYT Regular Grant 1202007.

\section*{Data Availability}
An introduction to {\gaea}, a list of our recent work, as well as
datafile containing published model predictions, can be found at
\url{http://adlibitum.oats.inaf.it/delucia/GAEA/}. In particular, the
predictions used in this paper will be shared on request to the
corresponding author. The latest VANDELS data release is available
through our public database at \url{http://vandels.inaf.it/dr3.html}
or via the ESO archive.

\bibliographystyle{mnras} \bibliography{fontanot}

\end{document}